# Microparticle laser fragmentation in liquids: mechanisms, energetics, and efficiency quantified with single-pulse, single-particle precision

Maximilian Spellauge[1,2], Ramon Auer[1], Meike Tack[2], Florentine Limani[1], David Redka[1,3], Sven Reichenberger[2], Anna R. Ziefuss[2], Stephan Barcikowski[2] and Heinz P. Huber[1,3]

[1]Laser Center HM, Munich University of Applied Sciences HM, 80335 Munich, Germany

[2]Technical Chemistry I and Center for Nanointegration Duisburg-Essen (CENIDE), University of Duisburg-Essen, 45141 Essen, Germany

[3]New Technologies Research Center, University of West Bohemia, Plzen CZ-30100, Czech Republic

## Abstract

Microparticle laser fragmentation in liquids has emerged as a promising approach to generate nanoparticles with high efficiency. Despite its advantages, the underlying fragmentation mechanisms, their connection to the nanoparticle size distribution, and the energy efficiency of the process remain poorly understood. In this study for the first time, microparticle fragmentation is investigated in single-pulse, single-particle experiments on Au microparticles. Determining the absorbed peak fluence enables assessment of the process energetics. Pump–probe microscopy identifies photomechanical fracture of the molten microparticle volume and photothermal phase explosion of its superheated surface as the fragmentation mechanisms. We find that 83% of the absorbed laser energy is converted into cavitation bubble formation, while only 1% contributes to the surface energy of the generated nanoparticles. Despite this small fraction, MP-LFL outperforms laser ablation in liquids. The surface energy generated per absorbed energy is 10 times higher, and the overall energy efficiency is 14 times higher. This gain originates from the confined microparticle geometry, which minimizes energy losses and enhances photomechanical fragmentation via pressure focusing. These results position microparticle fragmentation in liquids as a fundamentally more energy-efficient approach for scalable, laser-based nanoparticle production than laser ablation in liquids.

# 1. Introduction

Laser synthesis and processing of colloids (LSPC) represents a versatile and scalable technique for producing nanoparticles (NPs) [1,2]. Compared to conventional chemical synthesis routes, LSPC enables ligand-free NP production without precursor chemicals [3]. Additionally, LSPC can be applied to a broad range of materials, including metals, oxides, and semiconductors, making it an attractive method for advanced nanomaterial synthesis [1].

Within LSPC, laser ablation in liquids (LAL) [4] and its fragmentation-based derivatives, NP laser fragmentation in liquids (NP-LFL) [1,5] and microparticle laser fragmentation in liquids (MP-LFL) [6–10], represent key approaches for synthesizing colloidal NPs. LAL utilizes a solid bulk target immersed in liquid and enables NP generation in a single step with productivities up to 8.3 g/h [11], making it competitive with chemical synthesis routes [12]. However, it suffers from the need for continuous target replacement. NP-LFL, in contrast, enables continuous size reduction of pre-synthesized NPs, but depends on an additional synthesis step [13]. By combining these advantages, MP-LFL offers a scalable route that utilizes inexpensive feedstock suspensions, thereby avoiding both target replacement and the need for pre-synthesized NPs [9].

From a mechanistic perspective, MP-LFL was initially assumed to be dominated by photothermal mechanisms [8,14], similar to NP-LFL [15]. In LAL, laser pulse absorption leads to rapid heating of the particles with heating rates of several 1000 K/ps followed by surface evaporation and phase explosion [16,17]. Subsequent NP nucleation explains the formation of NPs with sizes below approximately 10 nm [16,17], whereas single-pulse NP-LFL fragmented in the phase explosion regime yields far smaller particles [15,18,19]. However, several experimental observations have challenged a purely photothermal model. Enhanced fragmentation efficiency was observed for MPs that had been mechanically treated by wet-grinding prior to irradiation, suggesting that introduced lattice defects promote photomechanical fracture [20]. In other experiments, fragmentation occurred even at laser fluences below the vaporization threshold, consistent with photomechanical fragmentation rather than thermally induced melting or evaporation [21]. Furthermore, the frequent occurrence of broad [8] or bimodal [9] size distributions indicates that both photothermal and photomechanical fragmentation mechanisms are active during LFL, and pump–probe microscopy (PPM) has revealed signatures of photomechanical fragmentation [9]. These findings motivated the hypothesis that stress confinement, which generates high tensile pressure inside the MP, promotes its fragmentation [9]. To date, however, it remains unresolved whether photothermal or photomechanical processes dominate MP-LFL, nor which NP sizes they predominantly generate.

Addressing this mechanistic uncertainty requires quantitative insight into the energetics of MP-LFL, yet prior studies have mostly focused on productivity benchmarks. In particular, MP-LFL has been shown to achieve high productivities in liquid jets, enabling continuous operation [7,22]. Additionally, it has been shown that MP-LFL robustly produces nanoclusters with sizes below 3 nm [23], further broadening its applicability. However, in most prior studies, MP-LFL efficiencies and size distributions are inferred from multiple interaction events, as multiple MPs are irradiated per pulse and individual MPs can be exposed to 20 to 400 pulses in a liquid jet reactor [23]. Moreover, in liquid jet or batch reactors, it is challenging to quantify the amount of energy absorbed by individual MPs, as absorption and scattering within the colloidal ensemble lead to fluence gradients and thus inhomogeneous irradiation [24]. In addition, absorption quantification can be complicated by the heterogeneity of the MPs themselves, as reported in [9], where the IrO$_2$ MPs used already exhibited irregular shapes, granular surfaces, and a broad size dispersion spanning several micrometers.

Without precise knowledge of the absorbed energy per particle, the fundamental energy requirement for fragmentation cannot be established. Such energetic information provides critical context for assessing which fragmentation mechanisms are plausible [9]. In addition, the fraction of absorbed energy converted into NP surface area directly links MP-LFL energetics to application relevance, since the generated surface area governs catalytic activity in heterogeneous catalysis [25].

This work systematically investigates MP-LFL in a well-defined approach where single Au MPs are irradiated by single laser pulses. This allows to determine the amount of laser energy absorbed by the MP. Furthermore, by fragmenting directly on a transmission electron microscopy (TEM) grid, the size distribution of the generated NPs can be determined without relying on post-fragmentation separation methods such as sedimentation or centrifugation. By observing the transient relative reflectance change during the process utilizing PPM [26] and quantifying the fragmentation energetics, the dominant fragmentation mechanisms are identified and linked to specific size fractions of the generated NPs. Finally, the energy efficiency of MP-LFL is compared to LAL, and its scalability is assessed.

## 2. Materials and methods

### 2.1 Microparticle characterization and sample preparation

Au MPs (99.9% purity; Evochem, charge number 19072301) were used as the starting material. As shown in Figure 1A, their ferret diameter distribution (Supplementary Information, Section S1) was determined using scanning electron microscopy (Apreo S LoVac, Thermo Fisher Scientific). Analysis of 139 MPs yielded an average ferret diameter of $D_{MP} = (1.2 \pm 0.2)$ µm. MPs were suspended in deionized water at a concentration of 500 mg/L and ultrasonically dispersed prior to use.

For fragmentation threshold determination (Section 3.1) and PPM (Section 3.2), the Au MPs were positioned on a polished sapphire substrate with a diameter of 50 mm and a thickness of 1 mm. A single droplet (20 µL) of the Au MP suspension was drop-cast onto the substrate and dried at 60°C. A second droplet (20 µL) of deionized water was applied to immerse the MPs before laser irradiation. To allow post-fragmentation analysis of the generated fragments (Section 3.3), a second experiment was conducted where the substrate was replaced with a TEM grid (EMR Lacey Carbon support film on a nickel 400 square mesh, diameter 2 mm). For this, graphene oxide (GO) nanosheets were prepared from a 1 g/L suspension in deionized water (stirred for 1 h and sonicated for 2 h to obtain thin GO nanosheets [27]) and deposited onto the TEM grid by drop-casting 20 µL, followed by drying at 60°C. The GO coating was used to fill the large, unsupported area of the TEM grid to capture the generated NPs. A droplet (20 µL) of the Au MP suspension was dropped onto the GO-coated TEM grid and dried at 60°C. Finally, a droplet (20 µL) of deionized water was applied to immerse the MPs, spanning the entire TEM grid surface.

### 2.2 Pump-probe microscopy setup

A femtosecond laser (Spirit 16W HE SHG F2P CS, Spectra Physics) served as the PPM laser source. The laser operated at 120 µJ pulse energy, 1040 nm wavelength, 500 fs pulse duration, and 500 Hz repetition rate. A half-wave plate (HWP) combined with a polarizing beam splitter (PBS) split the pulses into pump and probe branches with an energy ratio of 9:1. Mechanical shutters selected single pulses in each branch to allow for single-pulse fragmentation of the MP (pump) and single-pulse imaging of the reflectance change (probe).

In the pump branch, the pump-pulse energy was adjusted to a few 100 nJ using a second HWP-PBS combination. A pulse stretcher increased the pump pulse duration to 10 ps (FWHM, measured by autocorrelation assuming a Gaussian temporal intensity distribution) [28].

In the probe branch, a second-harmonic generation unit converted the wavelength to 520 nm. The probe pulse was optically delayed relative to the pump pulse, enabling delay times $\Delta t$ ranging from -100 ps to 5 ns. For delays up to the millisecond range, a second electronically delayed laser (piccolo AOT, Innolas Laser GmbH) with a wavelength of 532 nm and a pulse duration of 500 fs was used.

The pump and probe branches were colinearly combined in the microscopy setup (Figure 1B). A dichroic mirror reflected the pump pulse into a water-immersion objective (Leica, HC Fluotar, 25x, NA = 0.95). The objective focused the linearly polarized pulse through the 0.6 mm water gap, a 150 μm thick cover glass positioned 2 mm above the sample, and the water layer beneath it onto the substrate (Figure 1C). The beam waist radius was determined as $w_0 = (2.6 \pm 0.1)$ μm (Supplementary Information, Section S2), and the incident peak fluence $\Phi_I$ was calculated by $\Phi_I = 2 \cdot T \cdot E_0 / (\pi \cdot w_0^2)$. Here $E_0$ denotes the pulse energy measured after the objective in the absence of the water layer, and $T = 0.95$ is the linear transmittance through the 2.6 mm water layer (Figure 1C) [29]. At 10 ps pulse duration, the critical peak fluence for optical breakdown is 15 J/cm² [9], so interaction with the water layer can be assumed to be linear. Reflectance losses at the cover glass are < 1% and are neglected. Given 1% pulse-to-pulse energy fluctuations and the uncertainty in the beam waist radius, the relative uncertainty of the peak fluence is approximately 8%.

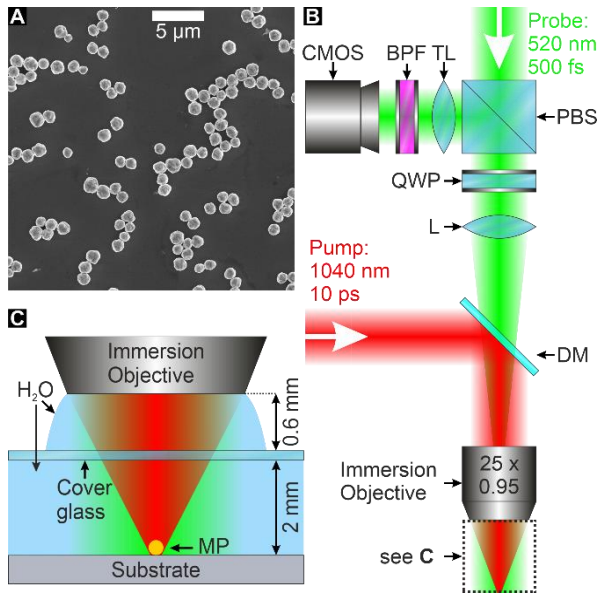

**Figure 1:** Microparticle characterization and laser fragmentation in liquids setup. **A** SEM image of Au microparticles used as the starting material for the fragmentation experiments. **B** Optical setup for the fragmentation experiments, including the key components: polarizing beam splitter (PBS), quarter-wave plate (QWP), lens (L), dichroic mirror (DM), tube lens (TL), and bandpass filter (BPF). Laser pulses with a wavelength of 520 nm (green) and 1040 nm (red) were colinearly guided onto the sample. **C** Close-up of how the microparticles were immersed in water and illuminated by the laser source.

The probe imaging path followed the established PPM configuration [30–32]. It was circularly polarized and had a peak fluence of approximately 1 mJ/cm². The pump pulse was imaged onto a CMOS camera (pco.pixelfly usb, PCO AG) via a tube lens, providing a lateral resolution of approximately 0.5 μm. A $(520 \pm 10)$ nm band-pass filter in front of the camera suppressed scattered pump and ambient light. The camera and both mechanical shutters were temporally synchronized by a digital delay generator (DG645, Stanford Research Systems), with the ultrafast laser source serving as the master trigger

At each delay time, a single pump pulse irradiated a single MP, and the image $R(\Delta t)$ was recorded. A reference image $R_0$ was acquired 1 s before irradiation. The procedure was repeated for each subsequent delay by manually selecting a previously unirradiated MP. The relative reflectance change was computed pixel-wise by $\Delta R/R_0 = (R(\Delta t) - R_0)/R_0$. Zero delay ($\Delta t = 0$ ps) was defined as the time at which the pump and probe intensity maxima overlap and was determined from the instantaneous reflectance increase of an indium tin oxide thin film under sub-threshold pumping [33]. All measurements were performed at room temperature $(25 \pm 1)$ °C

## 2.3 Post-fragmentation nanoparticle characterization

A single Au MP on the graphene oxide-coated TEM grid was located with the microscopy setup (Figure 1B) and fragmented with a single pump pulse. After irradiation, the grid was removed and dried at ambient conditions. The fragmentation site was relocated by TEM (Probe Side CS-corrected JEM-2200FS, JEOL, accelerating voltage of 200 kV). The surrounding region of 78 μm x 78 μm, corresponding to approximately 2% of the total water-covered TEM grid area, was then examined. For NP statistics, six TEM images were acquired (Supplementary Information, Section S3). The NP feret diameters were measured using the software ImageJ.

## 3. Results

### 3.1 Microparticle fragmentation threshold

The absorbed peak fluence was calculated as $\Phi_{abs} = A_{MP} \cdot \Phi_I$, where the MP absorptance $A_{MP}$ was obtained from Mie theory using a fluence-dependent extinction coefficient derived from bulk Au data (Supplementary Information, Section S4). The resulting absorptance is $\approx 5.8\%$ up to an incident peak fluence of $\Phi_I = 1.5$ J/cm² and increases thereafter.

The single-pulse fragmentation threshold $\Phi_{thr}$ was determined by irradiating individual Au MPs with increasing absorbed peak fluence. For each fluence, ten MPs were irradiated, and two images were recorded per MP (Figure 2A). At 30 mJ/cm², the image at $\Delta t = 4$ ns shows only a slight reflectance increase, and that at $\Delta t = 1$ s shows the MP intact. At 50 mJ/cm², the 4 ns image reveals a shockwave [34] and cavitation bubble [9,31,35], while the 1 s image shows diffraction fringes indicating the generation of fragments.

The transition from no observable fragmentation at 30 mJ/cm² to complete fragmentation at 50 mJ/cm² was gradual rather than abrupt. For intermediate fluences, some MPs remained intact while others fragmented, indicating probabilistic behavior. Therefore, the fragmentation probability $P_{frag} = N_{frag}/N_0$ was defined, where $N_{frag}$ and $N_0 = 10$ denote the number of fragmented and irradiated MPs, respectively. As shown in Figure 2B, $P_{frag}$ rises from 0% to 100% between 30 mJ/cm² and 46 mJ/cm².

The random nature of the MP-LFL process is attributed to particle-to-particle variations in size, surface morphology or internal stress, since uncertainties from pulse-to-pulse fluctuations and size-dependent absorptance were below 2 mJ/cm² and thus cannot explain the broad transition region. The transition region is well described by an error function, consistent with a random process. The fragmentation threshold $\Phi_{\text{thr}}$, defined at 50 % probability, corresponds to an absorbed peak fluence of 37 mJ/cm$^2$ and an incident peak fluence of 640 mJ/cm$^2$.

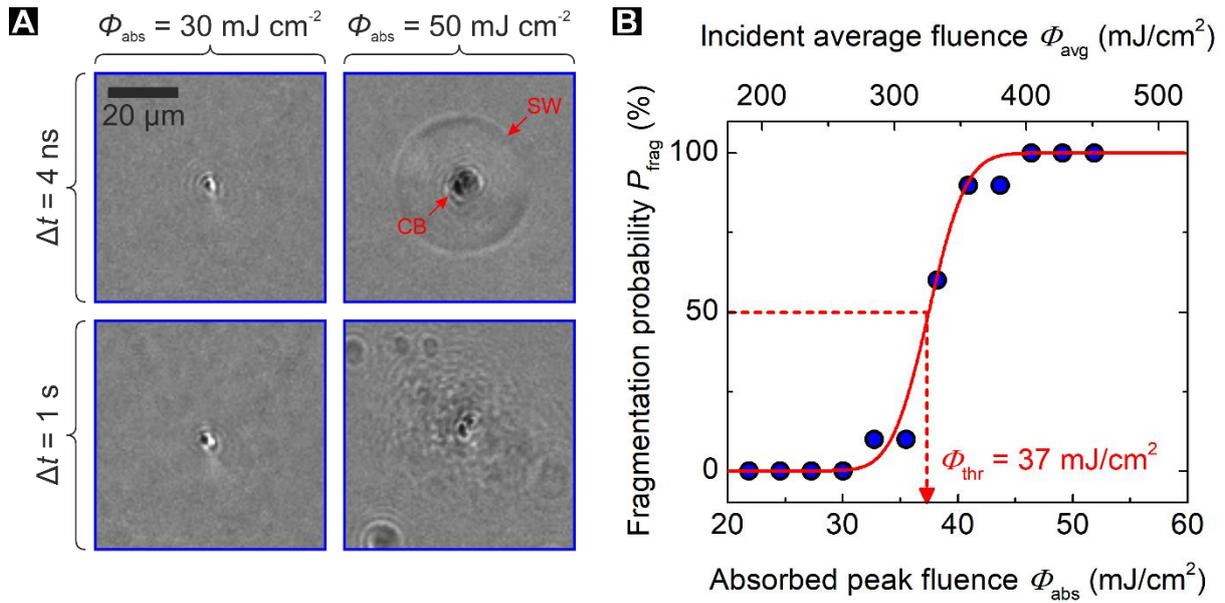

**Figure 2:** Single-pulse fragmentation threshold of Au microparticles. **A** Time-resolved microscopy images of Au microparticles immobilized on a sapphire substrate and irradiated in water using single laser pulses with a pulse duration of 10 ps. Different absorbed peak fluences $\Phi_{\text{abs}}$ are displayed for delay times of 4 ns (top row) and 1 s (bottom row) **B** Fragmentation probability $P_{\text{frag}}$ as a function of $\Phi_{\text{abs}}$. The fragmentation threshold of $\Phi_{\text{thr}} = 37$ mJ/cm$^2$ is defined at a fragmentation probability of 50% and marked by a red dashed vertical line. For comparison, the incident average fluence $\Phi_{\text{avg}}$ is given on the top axis.

At absorbed peak fluences above 46 mJ/cm$^2$ (corresponding to an incident peak fluence of 800 mJ/cm$^2$), fragmentation occurs with 100% probability, ensuring reliable single-pulse fragmentation of individual MPs. These findings establish a clear peak fluence range for MP-LFL, which serves as the foundation for the investigation of MP-LFL dynamics in the next section.

## 3.2 Microparticle fragmentation dynamics

Figure 3A shows the spatio-temporal reflectance change $\Delta R/R_0$ following single-pulse irradiation of an Au MP at an absorbed peak fluence of 560 mJ/cm$^2$ (incident fluence of 2850 mJ/cm$^2$). Videos of the full image sequence are provided in the Supplementary Material. This fluence lies well within the phase-explosion regime, corresponding to about 15 times the fragmentation threshold and 3.5 times the phase-explosion threshold of 160 mJ/cm$^2$ (see Supplementary Information, Section S5), and was chosen to test whether photomechanical effects contribute to fragmentation in this regime.

At $\Delta t = 0$ ps, a drop in $\Delta R/R_0$ appears at the MP position, attributed to electron-plasma formation above the surface via thermionic emission, as observed for ablation of Au in water at similar fluences [31]. By $\Delta t = 100$ ps, a radially expanding ring emerges, which is visible up to 10 ns and corresponds to a shockwave that propagates in water [9].

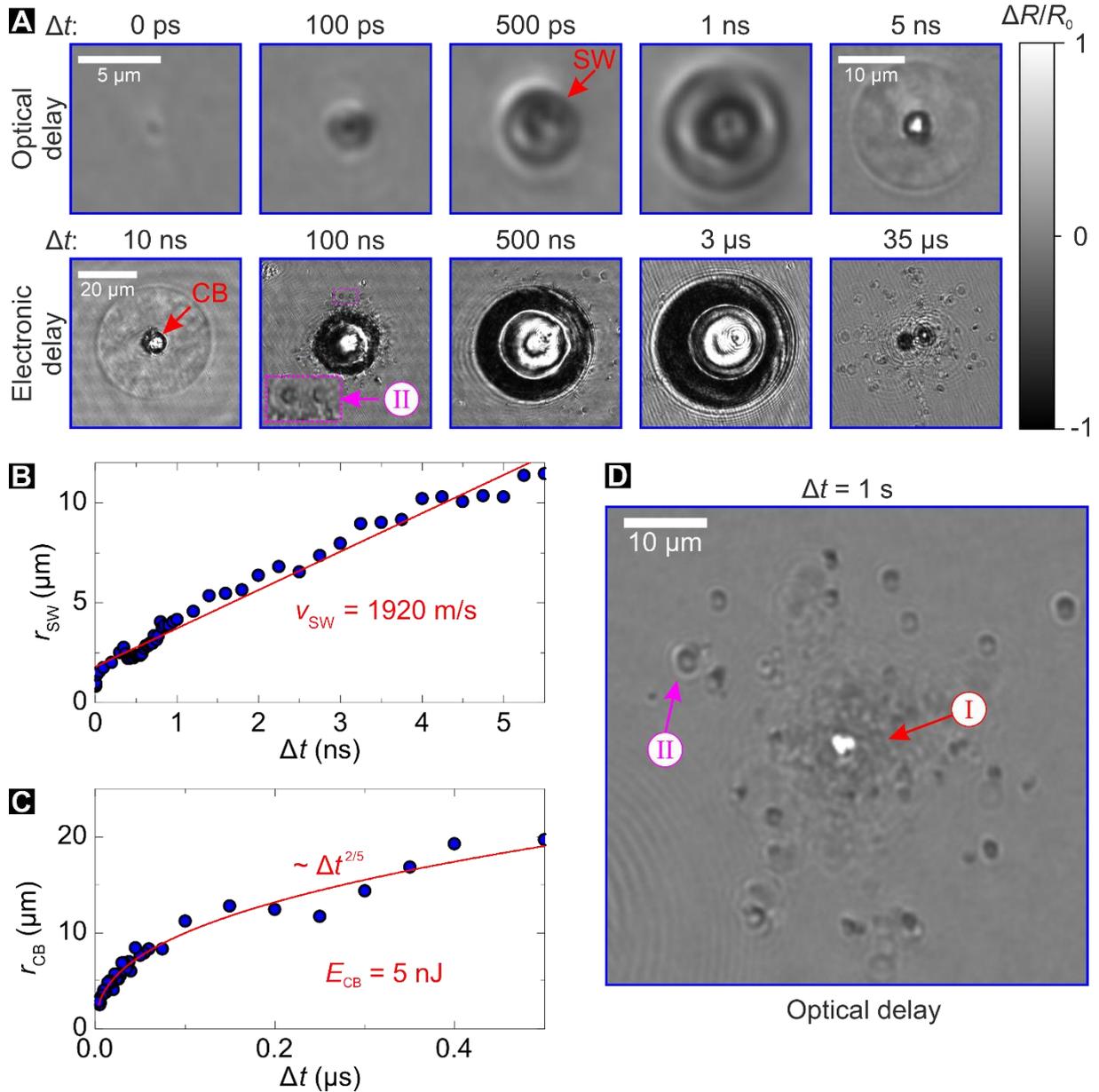

**Figure 3:** Fragmentation dynamics of single Au microparticles irradiated with a pulse duration of 10 ps and an absorbed peak fluence of $\Phi_{abs} = 560$ mJ/cm$^2$. **A** Images depicting the relative reflectance change $\Delta R/R_0$ for different delay times $\Delta t$. Negative values correspond to decreased reflectance (dark), positive values to increased reflectance (bright). Each scale bar applies to all following delay times within the same row until another scale bar appears. The inset marked II at $\Delta t = 100$ ns corresponds to an enlarged view of the diffraction patterns indicated by a magenta dotted frame. This diffraction pattern is speculated to originate from generated nanoparticles that are larger than 100 nm. The shockwave and cavitation bubble are indicated by red arrows and labeled SW and CB, respectively. **B** Temporal evolution of the shockwave radius $r_{SW}$ with linear fit. **C** Temporal evolution of the cavitation bubble radius $r_{CB}$ with $\Delta t^{2/5}$ fit. **D** $\Delta R/R_0$ at $\Delta t = 1$ s. Nanoparticles smaller than approximately 30 nm are marked with I and nanoparticles larger than approximately 100 nm are marked with II.

Figure 3B shows the temporal evolution of the shockwave radius $r_{SW}$. A linear fit yields a velocity of 1920 m/s, about 30% higher than the sound velocity in water (1480 m/s [36]).

At $\Delta t = 1$ ns, a region of decreased reflectance forms around the MP, indicating cavitation-bubble formation (Figure 3A) [9]. At 5 ns, a bright central spot appears, caused by focusing of the probe beam at the curved bubble interface [37]. The bubble expands from the nanosecond to microsecond timescale, reaching a maximum radius $R_{max} = (23 \pm 1)$ μm at $\Delta t \approx 3$ μs (Figure 3A). From this maximum, the cavitation-bubble energy is estimated as $(5.0 \pm 0.7)$ nJ (Supplementary Information, Section S6) [38].

The bubble energy was also estimated from its expansion dynamics (Figure 3C) [34], yielding $(5.0 \pm 0.5)$ nJ (Supplementary Information, Section S7). This value is in excellent agreement with the value derived from the maximum radius, demonstrating the consistency of both methods.

After reaching its maximum expansion at $\Delta t = 3$ μs, the cavitation bubble collapses by 35 μs (Figure 3A). At this time, a dark region of reduced reflectance remains near the initial MP position, corresponding to a long-lived microbubble surrounded by fragments. This microbubble disappears at a delay time of 1 s (Figure 3D).

A closer examination of the image recorded at 1 s (Figure 3D) reveals two distinct types of fragments.

Fraction I appears as a dark, diffuse cloud surrounding the initial MP position and is characterized by a local reflectance decrease (I in Figure 3D). The absence of clear interference fringes suggests that scattering from these NPs is negligible and that the observed signal originates primarily from absorption [39,40], pointing to NPs smaller than 30 nm, for which scattering is negligible (Supplementary Information, Section S8). Importantly, this size range is still above the single-particle detection limit of our PPM setup and therefore detectable. The detection limit was estimated by using the criterion that the absorption produced by single NP located within the point-spread function area of the PPM must generate a $\Delta R/R_0$ change that exceeds the 2% noise level. By this the smallest detectable NP diameter was estimated to be approximately 26 nm (Supplementary Information, Section S9). In addition, ensembles of even smaller NPs contribute additively to absorption [41], so that ensembles of sub-26 nm NPs are also expected to be detectable. The spatial localization of fraction I within a radius of 10 μm of the initial MP position further supports this assignment, consistent with simulations predicting that small NPs (< 10 nm) remain close to the fragmentation site [18,42] and may even rebound at the cavitation bubble due to the inverse Leidenfrost effect [15].

Fraction II consists of discrete fragments at distances of several tens of μm from the initial MP that exhibit circular interference fringes (II in Figure 3D). The presence of such fringes indicates that scattering is substantial, implying NP sizes larger than 100 nm (Supplementary Information, Section S8). Fraction II is already visible beyond the cavitation bubble boundary at 100 ns (II in Figure 3A) and appears outside the maximum cavitation bubble radius in the final state (compare $\Delta t = 3$ μs in Figure 3A and $\Delta t = 1$ s in Figure 3D). This is consistent with computational [17] and experimental [17,43,44] findings that NPs with diameters of several tens of nanometers are ejected beyond the cavitation bubble boundary.

Taken together, these observations support the presence of two distinct NP size fractions after MP-LFL. A quantitative assessment of the NP size distribution based on TEM is presented in the next section.

### 3.3 Nanoparticle analysis following single microparticle fragmentation

NPs generated by irradiating a single MP supported on a GO-coated TEM grid were examined by TEM. The MP was irradiated at the same absorbed peak fluence of 560 mJ/cm$^2$ as in Section 3.2. Figure 4A shows an overview of the analyzed area, where NPs were detected within 10 μm of the original MP position (distance between orange crosshair and dashed line).

A total of 580 NPs were identified from six representative TEM images (see Supplementary Information, Section S3). Figure 4B depicts two representative TEM images of the generated NPs.

Size-distribution analysis (Figures 4C, 4D, and 4E) shows that the generated NPs fall into two distinct fractions. Fraction I, extending up to $\approx 72$ nm, is most abundant, comprising 98.6% of all detected NPs but contributing only 65% of the total NP surface area and 36 % of the total NP mass. Fraction II, spanning from 85 nm to 150 nm, represents just 1.4% of all NPs yet dominates the mass balance, accounting for 64 % of the total mass. The identification of two size fractions with the optical analysis of the final state, where two size fractions were also identified (Figure 3D). A similar size distribution was also reported for $IrO_2$ MP-LFL, where a narrow fraction centered at 3 nm coexisted with a broader one extending from several hundred nanometres to a few micrometres [9].

Analysis of the volume-weighted size distribution ($dV/dD_F$ in Figure 4D) and total NP volume shows that $\approx 1.5\%$ of the initial MP volume was recovered in the analyzed region. This represents a lower limit for the true amount of generated NP, as rupture of the carbon support film near the fragmentation site (* in Figure 4A) likely allowed some NPs to escape through the openings. Additionally, during post-fragmentation drying, NPs could diffuse across the droplet and thus the entire TEM grid. Since only 2 % of the TEM grid was imaged, the analyzed NP fraction is not representative of the total amount of generated NPs.

A further limitation of the TEM-derived size distribution arises from post-fragmentation diffusion. During drying, 5 nm NPs diffuse 5.5 times farther than 150 nm NPs (Supplementary Information, Section S10), making small NPs more likely to move beyond the imaged region. TEM also struggles to resolve particles below 5 nm, as low contrast and overlap with larger NPs reduce visibility [45]. Consequently, the size distribution is assumed to underrepresent small NPs and to be skewed toward larger NPs.

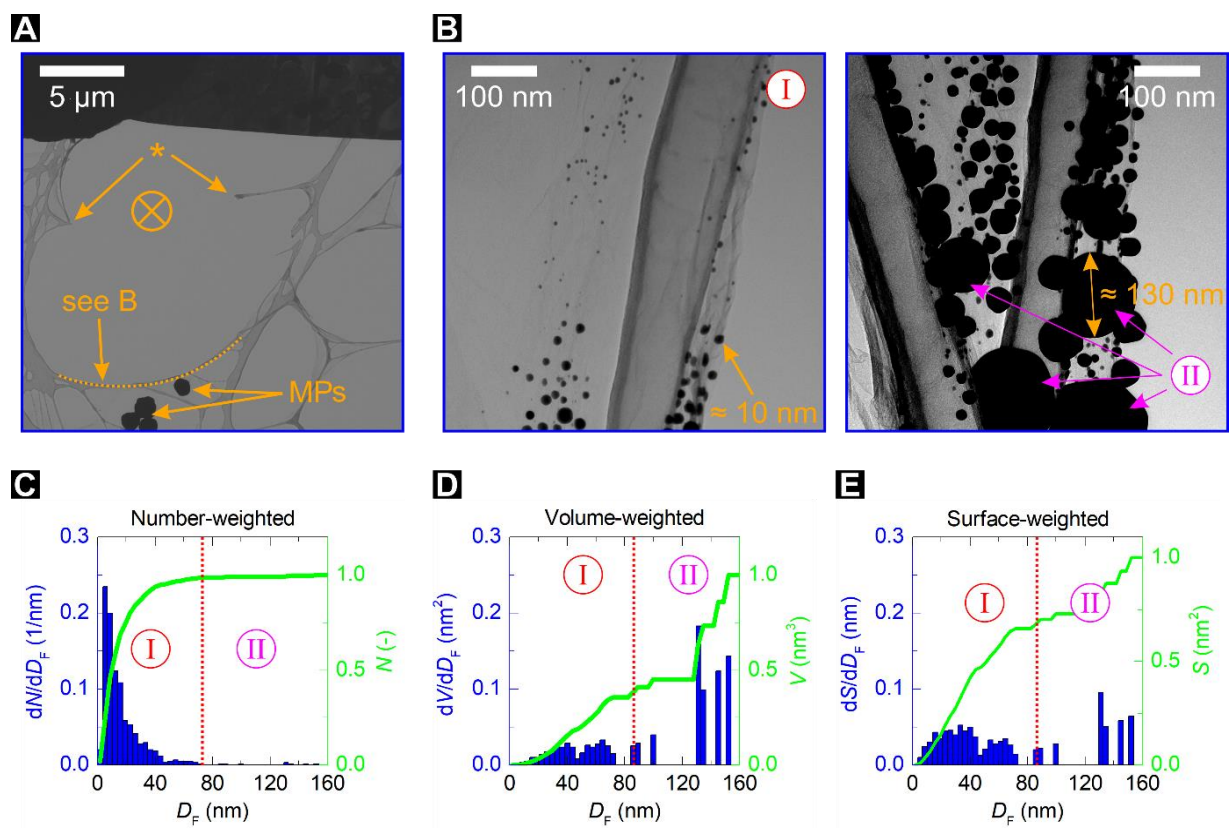

**Figure 4:** Nanoparticle size distribution obtained by fragmentation of a single Au microparticle. The microparticle was irradiated with a pulse duration of 10 ps and an absorbed peak fluence of $\Phi_{abs} = 560$ mJ/cm$^2$. **A** Overview of the region in the vicinity of the initial microparticle. The microparticle was situated at the position marked with an orange crosshair, and the NPs generated as a result of the fragmentation of the microparticle were found at the position marked with a dashed orange line. Positions marked by an asterisk show rupture of the carbon support film **B** Representative transmission electron microscopy images of nanoparticles detected within the orange dashed line in **A**. **C** Number-weighted NP size distribution obtained from 580 detected NPs across six TEM images. The normalized number-weighted size distribution d$N$/d$D_F$ is shown by blue bars and corresponds to the left y-axis, while the accumulated NP count is displayed by a green line corresponding to the right y-axis. **D** Normalized volume-weighted size distribution d$V$/d$D_F$. **E** Normalized surface-weighted NP size distribution d$S$/d$D_F$. **D** and **E** are structured identically to **C**. Two distinct size fractions (I and II) are identified and highlighted in **B**, **C**, **D** and **E**.

## 4. Discussion

### 4.1 Fragmentation threshold

To understand the energetic conditions under which MP-LFL initiates material removal, the measured fragmentation threshold fluence is first compared to literature values and to the ablation threshold of bulk Au in air and water. The absorbed fragmentation threshold fluence of 37 mJ/cm$^2$ determined in Section 3.1 is close to the 50 mJ/cm$^2$ reported for IrO$_2$ MP-LFL [9] and falls within the broad range of ultrafast NP-LFL thresholds reported in the literature, ranging from 1.3 mJ/cm$^2$ [46] to 187 mJ/cm$^2$ [13,47] (Supplementary Information, Section S11). Molecular-dynamics simulations predicting complete Au NP fragmentation at 36 mJ/cm$^2$ [18] corroborate this value. For comparison, the absorbed ablation thresholds of Au under 10 ps irradiation are approximately 100 mJ/cm$^2$ in air and 200 mJ/cm$^2$ in water [48]. Thus, the energetic requirement for the onset of MP-LFL is reduced by a factor of about 2.7 compared to ablation in air and by a factor of about 5.4 compared to ablation in water.

To place the threshold reduction for MP-LFL compared to LAL into context, the absorbed threshold fluence required to heat Au to its melting or vaporization energy density is estimated. For ultrafast excitation of metals, the laser energy is initially deposited within an effective penetration depth $d_{eff}$ [49]. The absorbed threshold fluence for melting or vaporization is then given by $\Phi_{thr} = u \cdot d_{eff}$ [50], where $u$ denotes either the volumetric energy density for melting $u_m$ or for vaporization $u_v$. With $d_{eff} \approx 100$ nm for Au [49], $u_m \approx 3.8$ J/mm$^3$ and $u_v \approx 41.7$ J/mm$^3$ (Supplementary Information, Section S12) the threshold fluences for melting and vaporization are estimated as 38 mJ/cm$^2$ and 417 mJ/cm$^2$, respectively.

These threshold estimates show that the experimentally measured MP-LFL threshold (37 mJ/cm$^2$) essentially coincides with the absorbed fluence required to reach the energy density for melting (38 mJ/cm$^2$), whereas the absorbed LAL threshold of Au in water (200 mJ/cm$^2$) is approximately half the absorbed fluence required to reach the energy density for vaporization (417 mJ/cm$^2$). Thus, MP-LFL initiates fragmentation at the energy density required for melting, while LAL requires energy densities more closely to vaporization, showing that MP-LFL is energetically more efficient than LAL.

While the threshold analysis identifies the energetic conditions required to initiate fragmentation, it does not quantify how much energy is actually required for the generation of

NPs. For this, the absorbed energy per unit volume required for NP generation in MP-LFL must be quantified. Asessing this quantity requires knowledge of the fragmentation yield, i.e. the fraction of the MP that is converted into NPs. However, the fragmentation yield could not be reliably determined from TEM analysis in Section 3.3. Therefore, to quantify the fragmentation yield and the energetics of MP-LFL, the fragmentation mechanisms are discussed in Section 4.2. Understanding the fragmentation mechanisms then enables an estimation of the NP yield in Section 4.3, which then allows to quantify the MP-LFL energetics in Section 4.4.

## 4.2 Fragmentation mechanisms

Before discussing the individual fragmentation mechanisms for an absorbed peak fluence of 560 mJ/cm$^2$ depicted in Figure 5, some assumptions are outlined. Mie simulations for 1.2 μm Ag MPs show that 1064 nm light is absorbed predominantly at the laser-facing hemisphere [51]. By analogy, for Au MPs at 1040 nm the effective absorption region is a 12.2 nm-deep [52] hemispherical shell (see panel (1) in Figure 5). With the beam waist radius of 2.6 μm greatly exceeding the MP radius of 0.6 μm, the local fluence across the MP is effectively uniform (deviation < 10 %). Furthermore, TEM shows nearly spherical fragments (Figure 4B), indicating fragmentation from the liquid phase, since solid-state fracture would yield irregular shapes [53]. In the following, we therefore examine the fragmentation mechanisms under these assumptions.

The melt depth of Au under conditions where the pulse duration (10 ps in our case) is shorter than the electron–phonon coupling time of 20 ps [54], is estimated as $\approx$ 360 nm at threshold (Supplementary Information, Section S13), corresponding to 60% of the 600 nm MP radius or $\approx$ 47% of its volume (Figure 5, panel 2). This agrees with simulations reporting a 450 nm melt depth for Au irradiated at 550 mJ/cm$^2$ [55]. The molten layer persists for $\approx$ 3 ns [55], indicating that under the present conditions 47% of the MP remains molten up to the nanosecond timescale. Because the melt depth was evaluated at threshold, the molten volume fraction of 47% represents a conservative estsimate. At the absorbed peak fluence of 560 mJ/cm$^2$ (approximately 15 times the threshold), the molten volume is expected to be larger.

The ultrashort-pulse ablation mechanisms, photothermal phase explosion and photomechanical spallation, are well known from laser ablation in air and in liquids [31,33,56]. We therefore assess whether these mechanisms also govern MP-LFL and whether the microparticle geometry alters these mechanisms.

*Phase explosion* may occur when the material is heated up to 90% of its critical temperature [57]. According to two-temperature model calculations (Supplementary Information, Section S5), this condition is reached for absorbed fluences above 160 mJ/cm$^2$. In this study, the absorbed peak fluence of 560 mJ/cm$^2$ exceeds that threshold by a factor of $\approx$ 3.5, indicating that phase explosion largely contributes to MP-LFL. Two-temperature hydrodynamic simulations of Au LAL in water support this interpretation, showing predominantly phase explosion at 550 mJ/cm$^2$, developing within a few picoseconds and affecting a $\approx$ 30 nm surface layer [58]. Thus, only a narrow surface region undergoes phase explosion (Figure 5, panel 3).

*Photomechanical fracture* requires that the stress-confinement condition is satisfied. This means that the time scale for heating, given by the longer of the pulse duration $\tau_P$ and the electron–phonon coupling time $\tau_{ep}$, must be shorter than the time scale for energy dissipation by an acoustic wave, given by the acoustic relaxation time $\tau_{ac}$. This condition is expressed as max$\{\tau_P, \tau_{ep}\} \leq \tau_{ac}$ [56], where $\tau_{ac} \approx 50$ ps [31], $\tau_{ep} \approx 20$ ps [54], and $\tau_P = 10$ ps. Since the stress confinement criterion is fulfilled, heating is effectively isochoric [59], resulting in a maximum pressure buildup within the MP. Fracture occurs only when this pressure exceeds a critical value. For molten Au, the relevant fracture threshold is the tensile stress for void nucleation [60], which increases with temperature and thus with depth from 2.6 GPa close to the surface (approximately 100 nm) [54] to about 7.7 GPa at the melting depth of 360 nm (Supplementary Information, Section S14). For solid Au, the corresponding dynamic spall strength is $\approx 11.6$ GPa [54]. To evaluate whether these limits are reached, the internal pressure was estimated following Reference [9]. The shockwave pressure is $\approx 400$ MPa, and the pressure associated with early cavitation-bubble formation (at $\Delta t = 20$ ps, corresponding to the electron–phonon coupling time for Au [54]) is $\approx 70$ MPa. Accounting for the acoustic reflectance at the Au–water interface ($R_{ac} = 0.92$, Supplementary Information, Section S15) gives a resulting surface pressure of $P_0 \approx 4.5$ GPa.

The surface pressure of 4.5 GPa exceeds the fracture threshold of 2.6 GPa for the upper layer of the molten volume [54], making photomechanical fracture of this region conceivable. However, it remains below the higher threshold of 7.7 GPa at the melt depth of 360 nm. To evaluate whether photomechanical fracture at the melt depth of 360 nm is nonetheless possible, the pressure wave's propagation within the molten hemispherical shell must be considered. With a melt depth of 360 nm and a sound velocity of 2300 m/s in molten Au [61], the travel time of 0.16 ns is far shorter than the 3 ns solidification onset [55], so the wave interacts with the entire molten region. Over this sub-micrometre distance, attenuation is negligible [54]. The MP's spherical geometry focuses the inward-traveling wave, leading to pressure amplification [62]. The amplification is approximated by $P(z) = P_0 \cdot r_{MP}/(r_{MP} - z)$, assuming that the inward-propagating wave behaves like a spherical pressure wave whose amplitude scales inversely with propagation distance. This yields an amplified pressure of $\approx 11$ GPa at the melt depth, about 2.5 times the surface value. Thus, the liquid-fracture threshold is exceeded, implying photomechanical fragmentation of the entire molten hemispherical shell (Figure 5, panel 4).

This interpretation is further supported by the fivefold lower threshold of Au MP-LFL compared to bulk Au LAL (Section 3.1). Near threshold, photomechanical fracture dominates, consistent with prior experimental [31] and computational [16,58,63] LAL studies. We therefore attribute the reduced MP-LFL threshold partly to the finite geometry of the MP, which leads to inward stress focusing so that pressures that would be insufficient to fracture a semi-infinite target can exceed the fracture criterion inside an MP.

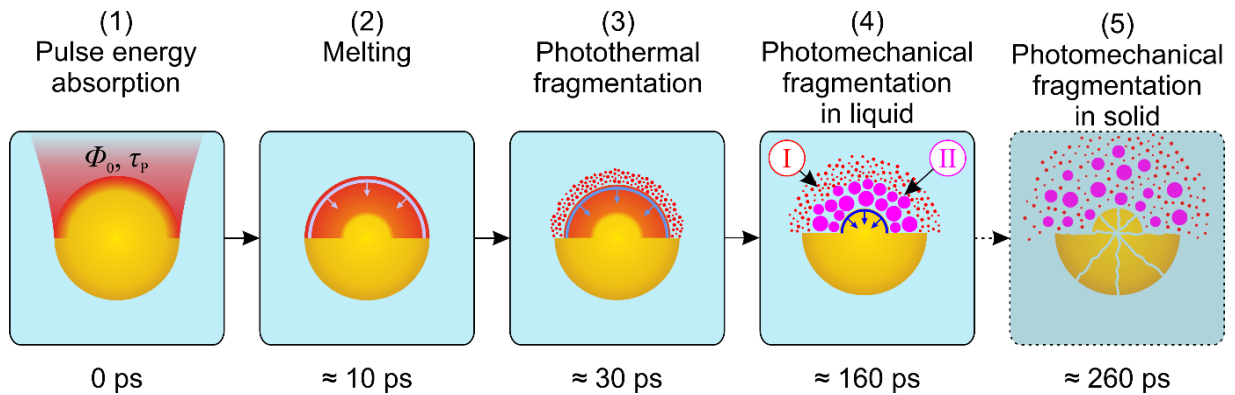

| (1) Pulse energy absorption | (2) Melting | (3) Photothermal fragmentation | (4) Photomechanical fragmentation in liquid | (5) Photomechanical fragmentation in solid |
|---|---|---|---|---|
| 0 ps | ≈ 10 ps | ≈ 30 ps | ≈ 160 ps | ≈ 260 ps |

**Figure 5:** Schematic depiction of microparticle fragmentation mechanisms at an absorbed peak fluence of 560 mJ/cm$^2$ and a pulse duration of 10 ps. (1) Pulse energy absorption within a hemispherical surface layer. (2) Formation of a molten hemispherical shell. (3) Photothermal phase explosion of a surface layer. (4) Photomechanical fragmentation of the molten volume, driven by an inward-propagating tensile wave (blue arrows). The pressure focusing effect is illustrated by the color transition from light to dark blue. (5) Speculative photomechanical fragmentation of the solid core. The two size fractions I and II identified in Figures 3 and 4 are marked.

The observation of two distinct size fractions, I and II, in both the TEM analysis (Figure 4, surface- and volume-weighted histograms) and the PPM experiments (Figure 3) further supports the identified fragmentation mechanisms. Fraction I ($\leq 72$ nm) agrees with established ultrashort-pulse LAL pathways [16,17,31,58,64]: NPs $< 10$ nm arise from vapor-phase condensation or droplet ejection during phase explosion [16,17,58], whereas tens-of-nanometre NPs result from hydrodynamic breakup of a molten surface layer released by photomechanical ablation [16,17]. The coexistence of these two surface processes is known to generate broad [17] or even bimodal [65] size distributions in LAL, spanning diameters up to 60 nm and 90 nm, respectively. Although no clear bimodality is observed in our case, Fraction I similarly spans a broad size range up to 72 nm, consistent with the simultaneous occurrence of photothermal phase explosion and photomechanical fragmentation of the surface layer. Fraction II (85 nm to 150 nm) cannot be explained by known LAL mechanisms and is instead attributed to photomechanical fragmentation of the molten hemispherical shell, promoted by pressure focusing in the spherical MP geometry. This additional contribution is unique to MP-LFL, distinguishing it from LAL, where such large fragments are absent [17,65].

Considering further pressure amplification deeper inside the MP, the pressure at the solid MP core may exceed the spallation threshold of 11.6 GPa for crystalline Au [54], making solid-state fracture conceivable (Figure 5, panel 5). For brittle IrO$_2$ MP-LFL, such fracture was reported, evidenced by faceted fragments larger than 100 nm [9]. In the present TEM analysis (Figure 4B), no faceted fragments are observed, only nearly spherical NPs. This difference is consistent with the higher brittleness of IrO$_2$ and the ductile nature of Au. Hence, while solid-core fracture cannot be excluded, a direct confirmation of solid-core breakup remains lacking. Taken together, these results show that both photothermal phase explosion and photomechanical fracture contribute to MP-LFL at an absorbed fluence of 560 mJ/cm$^2$ and 10 ps pulse duration, consistent with X-ray probing of Au MP-LFL [66]. Since the fragmentation threshold of 37 mJ/cm$^2$ (Figure 2) is about four times lower than the phase-explosion threshold of 160 mJ/cm$^2$ (Supplementary Information, Section S5), photomechanical fracture is expected to dominate near threshold, while the photothermal contribution increases with fluence [66].

## 4.3 Fragmentation yield

As the discussion of the fragmentation mechanisms in Section 4.2 showed, the molten hemispherical shell of the MP (approximately 47% of the MP volume) is expected to undergo fragmentation at the absorbed peak fluence of 560 mJ/cm$^2$. Assuming that the entire molten region is converted into NPs, this mechanistic picture gives an estimate of 47% for the fragmentation yield, i.e. the fraction of the MP mass converted into NPs.

The TEM-based estimate of 1.5% (Section 3.3) represents only a conservative lower limit, as NP diffusion during drying and rupture of the carbon support film caused substantial NP detection losses (Section 3.3). The mechanistic estimate therefore provides a more realistic picture of the actual yield and is used to assess the fragmentation energetics in Section 4.4.

However, it should be mentioned that the fragmentation yield of 47% contrasts with single-pulse, multi-particle MP-LFL experiments on non-spherical $IrO_2$ MPs in a liquid jet, which showed a yield of only 10% [9]. The lower value is expected to arise from material and size differences. In $IrO_2$, the absorbed energy is more strongly localized due to its lower thermal conductivity (100 W/m/K [67] versus 317 W/m/K for Au [68]) and its stronger electron–phonon coupling [67]. Consequently, only a smaller fraction of the MP volume is heated, and the large $IrO_2$ MP size of about 15 μm used in [9] further reduces the heated-volume fraction compared to the 1.2 μm sized Au MPs used in this study. At $\geq$ 20 pulses per MP, however, yields up to 50% were reported for Ir and $IrO_2$ [23], consistent with the 47% mechanistic estimate for Au under single-pulse conditions.

## 4.4 Fragmentation energetics and energy partitioning

To analyze the MP-LFL energetics, the distribution of the incident and the absorbed pulse energy is investigated, as shown in Figure 6. From the incident pulse energy of 303 nJ, the MP scatters 93 nJ (31%) and absorbs 6 nJ (2%) (Supplementary Information, Section S16). The absorbed portion of the laser energy is the energy available for fragmentation. As the energy of the cavitation bubble was found to be 5 nJ (see Section 3.2), it can be concluded that approximately 83% of the absorbed pulse energy is converted into cavitation bubble energy. Part of this fraction initially drives fragmentation and is subsequently released as heat, which expands the cavitation bubble. The determined cavitation bubble energy fraction agrees with MD simulations for 20 nm Au NPs [18], which found 80% of absorbed energy converted into bubble energy. Despite the different heating geometries (homogeneous heating of 20 nm NPs versus a hemispherical shell for 1.2 μm MPs), the results show that bubble formation dominates energy dissipation across sizes. Since the bubble is formed by heating the surrounding water, 80% of the absorbed energy ultimately heats the liquid. This is in contrast to LAL, where only 4% of the energy heats the liquid [69].

Assuming a fragmentation yield of 47%, the absorbed energy converted into NP surface energy is estimated as 0.06 nJ (Supplementary Information. Section S17), corresponding to $\approx$ 1% of the absorbed laser energy (Figure 6). Because TEM underrepresents the smallest NPs (Section 3.3), this value is a lower limit. To assess the possible contribution of undetected NPs, an upper-limit estimate assumes that the fragmented mass consists entirely of 5 nm NPs, yielding a total surface energy of 0.77 nJ, or $\approx$ 13% of the absorbed pulse energy.

Next, the fraction of laser energy converted into surface energy by MP-LFL is compared with that of LAL. The size distribution of Au NPs from LAL [17] under similar conditions (10 ps, 3400 mJ/$cm^2$ incident peak fluence) was analyzed. The total NP surface energy was estimated by weighting the detected NP volume of 2.4·$10^6$ $nm^3$ in [17] with the single-pulse ablation volume of 3.33·$10^9$ $nm^3$ reported in [48] for 10 ps and an absorbed peak fluence of 600 mJ/$cm^2$. This yields 1.24 nJ of surface energy for LAL of Au. With an absorbed energy of 1.1 μJ [48], only $\approx$ 0.1% of the absorbed energy is converted into NP surface energy, which is an order of magnitude lower than in MP-LFL, where $\geq$ 1% is converted. For context, conventional mechanical fragmentation by ball milling converts a fraction of 0.1% to 1% of the mechanical energy into surface energy [70]. Thus, MP-LFL reaches the upper limit of the efficiency range of ball milling while operating without the contamination issues associated with mechanical milling [10].

As can be seen in Figure 6, approximately 16% of the absorbed laser energy remains unassigned. Possible energy channels include residual heat in the MP or the surrounding water and the acoustic energy of the shockwave. A quantitative determination of the latter was not feasible, as it would require the shockwave pressure profile near its source [38] or a detailed intensity-dependent study [71]. However, Au LAL experiments report that 7% to 19% of the absorbed energy is converted into acoustic energy [71], suggesting that this component explains much of the unassigned fraction. In addition, chemical reactions during MP-LFL could release or consume energy, further affecting the balance [72].

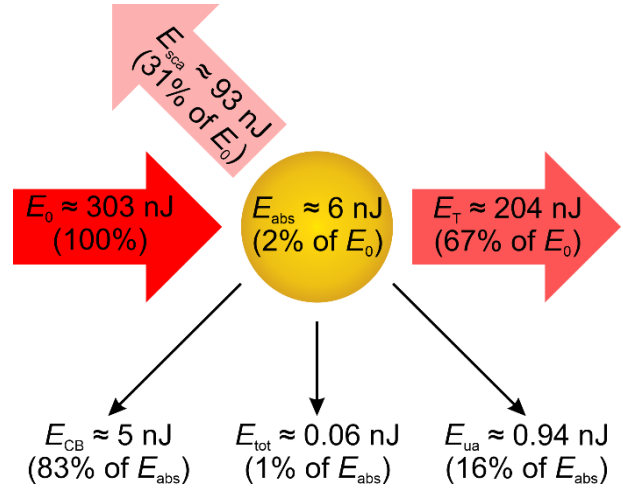

**Figure 6:** Energy balance of single-pulse, single-particle Au microparticle fragmentation. The incident pulse energy $E_0$ is partly scattered ($E_{sca}$), absorbed ($E_{abs}$), and unaffected by the MP ($E_T$) as indicated by red arrows. The absorbed energy is further partitioned into cavitation bubble energy $E_{CB}$, total surface energy of the generated nanoparticles $E_{tot}$, and unassigned energy $E_{ua}$, as indicated by black arrows.

We now consider the absorbed energy per unit volume required for NP generation in MP-LFL. This is approximately 14 J/mm³, based on 6 nJ of absorbed energy, a MP volume of 0.905 μm³ and a fragmentation yield of 47%. For comparison, melting and vaporization of Au require $\approx 3.8$ J/mm³ and $\approx 42$ J mm³, respectively (Supplementary Information, Section S12). Thus, MP-LFL requires approximately four times the energy to reach melting, but only around 33% of the energy required for complete vaporisation. Therefore, photothermal fracture alone cannot account for the fragmentation yield; photomechanical fracture must also play a significant role.

For comparison, Au ablation in water requires an absorbed energy density of $\approx 200$ J/mm³ to ablate the material at the most efficient peak fluence [48], about five times higher than the energy for complete vaporization. MP-LFL is therefore approximately 14 times more energy-efficient for generating Au NPs. This efficiency gain may be attributed to the confined geometry of the MP target. Unlike LAL, where approximately 34% of the absorbed laser energy dissipates into the bulk [73], MP-LFL confines the absorbed laser energy within the MP. In LAL, a significant portion of dissipated energy propagates as a pressure wave into the bulk. In MP-LFL, however, this pressure wave remains confined within the MP and may contribute to the fragmentation of the molten volume.

## 4.5. Comparison with state-of-the-art productivity benchmarks and assessment of productivity potential

While the absorbed energy defines the fundamental energetics of MP-LFL, process efficiency is better evaluated using the energy lost through absorption and scattering by the MP. This energy is estimated as 100 nJ (Figure 6). Assuming a 47 % fragmentation yield, this corresponds to 0.43 μm$^3$ (8.2 pg) of generated NPs per pulse. Dividing the energy lost through absorption and scattering by this volume yields a specific fragmentation energy of 235 J/mm$^3$, equivalent to a power-specific productivity of 300 mg/h/W.

For comparison, record Au LAL requires a specific energy of 3309 J/mm$^3$ [11], the fundamental Au LAL limit is 930 J/mm$^3$ [48], and 690 J/mm$^3$ is reported for IrO$_2$ MP-LFL [9]. This represents a substantial improvement: MP-LFL achieves the same material conversion with roughly 14-times less energy than record Au LAL, four-times less than the Au LAL limit, and three-times less than IrO$_2$ MP-LFL. Besides pressure focusing and energy confinement discussed in Section 4.3, this higher efficiency also stems in part from the approximately two-fold higher absorptance of Au MPs compared to bulk Au targets (Supporting Information, Section S4).

This promising result suggests that MP-LFL can enable a substantial increase in NP productivity. Extrapolating from the power-specific productivity of 300 mg/h/W and assuming an ultrafast laser source with 550 W average power [74], a productivity of up to 165 g/h could be envisioned. This value exceeds the record productivity of 4.7 g/h [11] achieved for LAL of Au under optimized conditions by a factor of 35. The experimental feasibility of reaching this value is evaluated in Section S18 of the Supplementary Information, where productivity is estimated under realistic continuous-flow flat-jet conditions. Section S19 discusses process limitations and scaling strategies to increase throughput, while Section S20 addresses approaches to narrow the NP size distribution.

## 5. Summary and conclusion

In this study, a single Au MP was fragmented by a single picosecond laser pulse. From the NP size distribution, energetic analysis, and time-resolved relative-reflectance change, the underlying fragmentation mechanisms were determined.

Two complementary fragmentation mechanisms were identified. Photothermal phase explosion vaporizes a surface layer of several tens of nanometers, whereas a photomechanical mechanism fractures the remaining molten hemispherical MP shell. The latter is supported by the tensile wave pressure exceeding the fracture threshold, enhanced by pressure amplification due to the nearly spherical geometry of the MP. This pressure amplification may also explain the fivefold lower fragmentation threshold observed for MP-LFL compared to LAL. Energetically, fragmentation occurs at roughly four times the melting requirement but only about one third of the vaporization requirement, which again suggests that the process is not purely photothermally driven but that photomechanical fracture contributes significantly. These mechanisms correlate with two characteristic NP size fractions, with mean diameters of approximately 10 nm and 120 nm, thereby explaining the observed broad NP size distribution.

Analysis of the process energetics further revealed that at least 1% of the absorbed pulse energy contributes to the surface energy of the generated NPs, while 83% is converted into cavitation bubble energy. The conversion of laser energy into NP surface energy is about an order of magnitude higher than reported for LAL. The incident energy requirement was determined to be 235 J/mm$^3$ for MP-LFL compared to 3310 J/mm$^3$ for record LAL, a factor of 14 higher energy efficiency. Based on this energy efficiency, MP-LFL productivity is extrapolated to 165 g/h, more than 30 times higher than the record productivity of LAL.

These results establish MP-LFL as a fundamentally more energy-efficient technique for laser-based NP production, surpassing the limits of LAL, while retaining its environmentally friendly character.

## Acknowledgements

The authors gratefully acknowledge financial support by the Deutsche Forschungsgemeinschaft (DFG) under projects 491072288, 428315411, 528706678 and 562785215. Anna Ziefuss further acknowledges funding by the DFG under project ID 405553726. Furthermore, we would like to thank Markus Heidelmann from ICAN, University of Duisburg-Essen, for the TEM measurements.

## S1. Microparticle educt size distribution from scanning electron microscopy analysis

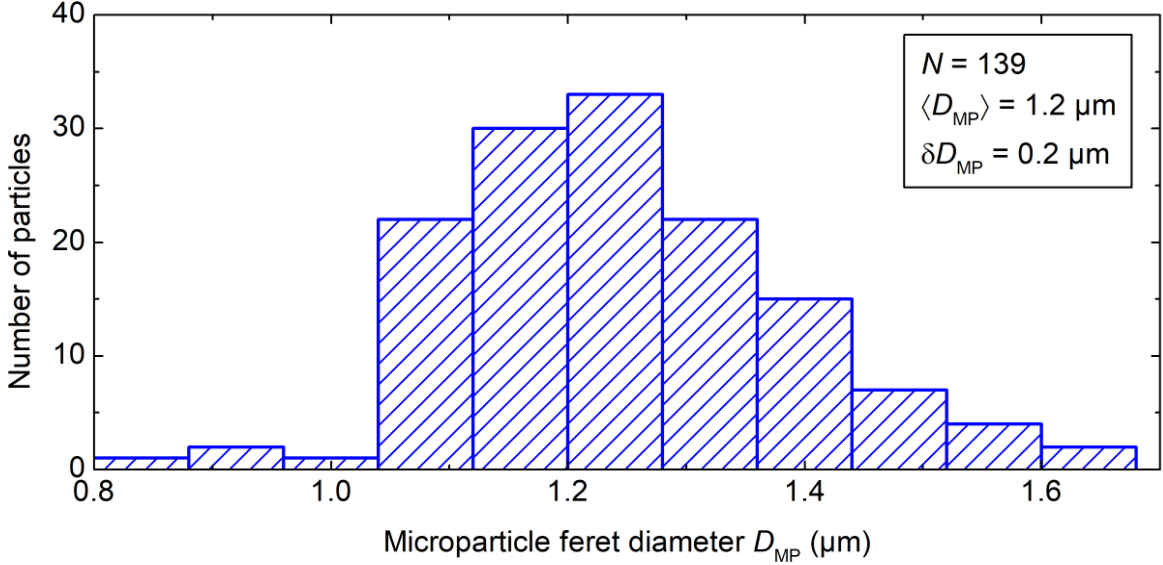

**Figure S1:** Histogram of feret diameters $D_{MP}$ for $N = 139$ microparticles analyzed from the SEM image shown in Figure 1A of the Main Manuscript. The mean microparticle diameter is $D_{MP} = (1.2 \pm 0.2)$ µm.

## S2. Determination of the beam waist radius

Assuming a Gaussian spatial intensity distribution, locality and an ideal threshold behavior, the ablation area $A_{abl}$ depends on the incident pulse energy $E_I$ (which accounts for the transmittance of 0.95 through a 2.6 mm thick water layer at 1040 nm) as given by Equation S1 [1].

$$A_{abl} = \frac{A_0}{2} \cdot \ln\left(\frac{E_I}{E_{thr}}\right) \tag{S1}$$

Here $E_{thr}$ denotes the ablation threshold pulse energy and $A_0 = \pi \cdot w_0^2$ denotes the beam waist area.

To determine the beam waist radius $w_0$ a silicon wafer was irradiated with increasing incident pulse energies $E_I$ using the pump-branch of the setup depicted in Figure 1 of the Main Manuscript. The ablation area was then measured using optical microscopy and is depicted in Figure S2. By fitting Equation S1 to the data (red line in Figure S2) a beam waist area of $A_0 = (21.2 \pm 0.8)$ µm² is obtained, resulting in a beam waist radius of $w_0 = (2.6 \pm 0.1)$ µm.

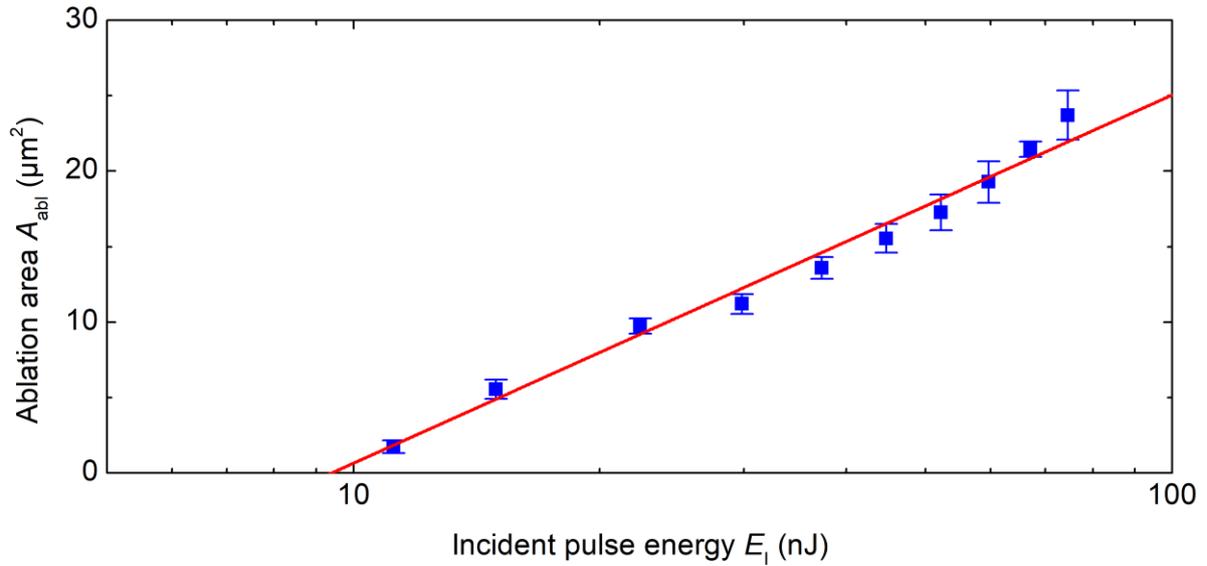

**Figure S2:** Ablation area $A_{abl}$ as a function of the incident pulse energy $E_I$. The ablation area was averaged over a total of 5 ablation craters obtained by irradiating a silicon wafer immersed in water with a wavelength of 1040 nm and a pulse duration of 10 ps. The red line is the fit of Equation S1 to the data points.

## S3. Transmission electron microscopy analysis of the generated nanoparticles

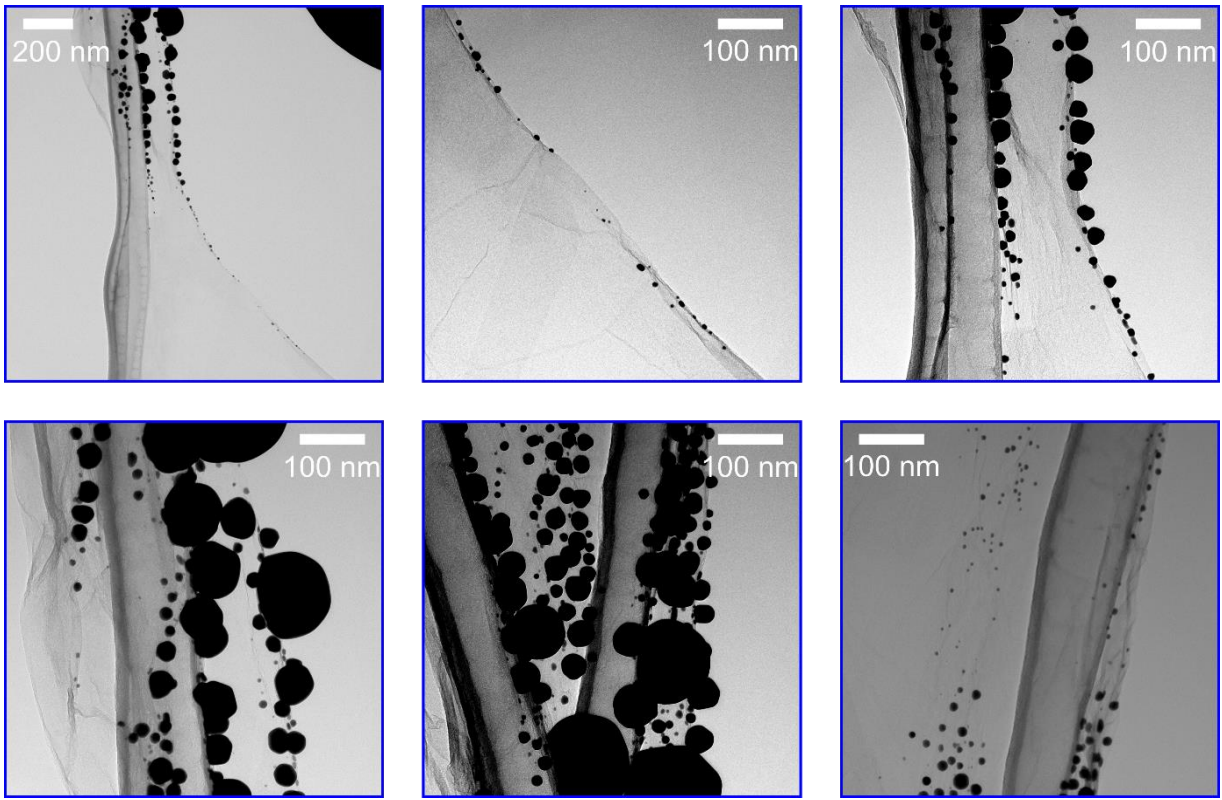

**Figure S3:** Transmission electron microscopy images of the nanoparticles generated following MP-LFL of Au MPs with an absorbed peak fluence of 560 mJ/cm$^2$.

## S4. Microparticle absorptance

Because the average microparticle (MP) diameter of $D_{MP} = 1.2$ μm (see Section S1) is comparable to the pump wavelength of 1.04 μm, optical absorption and scattering of the MP must be computed by Mie theory. This requires knowledge of the complex refractive index $N = n + i \cdot k$ of Au, where $n$ denotes the refractive index and $k$ the extinction coefficient. The complex refractive index of Au is highly dependent on the peak fluence [2]. Unfortunately, for a pulse duration of 10 ps, the fluence-dependent complex refractive index of Au is unknown. However, the fluence-dependent absorptance $A_{bulk}$ was previously measured on a polished bulk Au sample immersed in water for p-polarized light under an incidence angle of 25.6° as shown in Figure S4A [3].

It was observed that up to an incident peak fluence of $\Phi_l = 1.6$ J/cm$^2$, the absorptance remains constant at approximately 3%. To compare this constant value to the absorptance expected from literature data of the complex refractive index of Au, the reflectance $R$ is computed based on the Fresnel equation for p-polarized light under an incidence angle $\theta$ as given by Equation S2.

$$R = \left| \frac{-\left(\frac{N_2}{N_1}\right)^2 \cdot \cos\theta + \sqrt{\left(\frac{N_2}{N_1}\right)^2 - \sin^2\theta}}{\left(\frac{N_2}{N_1}\right)^2 \cdot \cos\theta + \sqrt{\left(\frac{N_2}{N_1}\right)^2 - \sin^2\theta}} \right|^2 \tag{S2}$$

Here $N_1$ denotes the complex refractive index of the target and $N_2$ the complex refractive index of the immersion medium. With $N_1 = 0.247 + i \cdot 6.781$ for Au at 1040 nm [4], $N_2 = 1.324 + i \cdot 1.63 \cdot 10^{-6}$ for water at 1040 nm [5] and $\theta = 25.6°$, Equation S2 yields a reflectance of $R = 0.97$, which given that the absorptance $A = 1 - R$ results in $A = 3\%$. As shown in Figure S4A the measured absorptance agrees well with the absorptance derived from literature values. Thus, for the low fluence regime ($\Phi_l \leq 1.6$ J/cm$^2$) the complex refractive index from [4] is used for Au as input for the Mie calculation.

As shown in Figure S4A, the absorptance increases approximately linearly when an incident peak fluence of $\Phi_l = 1.6$ J/cm$^2$ is exceeded, reaching 8% at $\Phi_l = 2.83$ J/cm$^2$. However, absorptance measurements alone cannot determine both the real part $n$ and imaginary part $k$ of the refractive index. A complete characterization would require fluence-dependent ellipsometry, which directly measures both components of the refractive index [6]. Nonetheless, pump-probe ellipsometry studies of Au thin films have shown that fluence dependent changes in the real part of the refractive index are negligible compared to variations in the imaginary part [7].

Based on this, a fluence-independent $n = 0.247$ for Au at 1040 nm [4] is assumed. The fluence-dependent imaginary part of the refractive index is estimated from the absorptance data shown in Figure S4A by iteratively varying $k$ in Equation S2 until the calculated absorptance matches the measured absorptance shown in Figure S4A. The resulting values for $k$ are depicted in Figure S4B and summarized in Table S1. For incident peak fluences up to 1.5 J/cm$^2$ the imaginary part $k$ remains constant at the literature value of 6.78 for 1040 nm [4]. Above this value $k$ decreases monotonically until it reaches 3.97 at an incident peak fluence of 2.85 J/cm$^2$.

Mie calculations were performed using MiePlot (version 4.6.21) [8]. In the experiments, the MP was located at the focal plane of the Gaussian beam. Here the wavefronts are planar and thus the calculation was performed for a plane wave at a wavelength of 1040 nm. The average MP diameter was set to $\langle D_{MP} \rangle = 1.2$ μm, and the real part of the refractive index $n = 0.247$ [4] was assumed to be fluence-independent, while the fluence-dependency was accounted for by using the imaginary part of the refractive index depicted in Figure S4B. Water, with a refractive index of $n = 1.324$ at 1040 nm [5] was the immersion medium. The results of the Mie calculation assuming spherical MPs are summarized in Table S1.

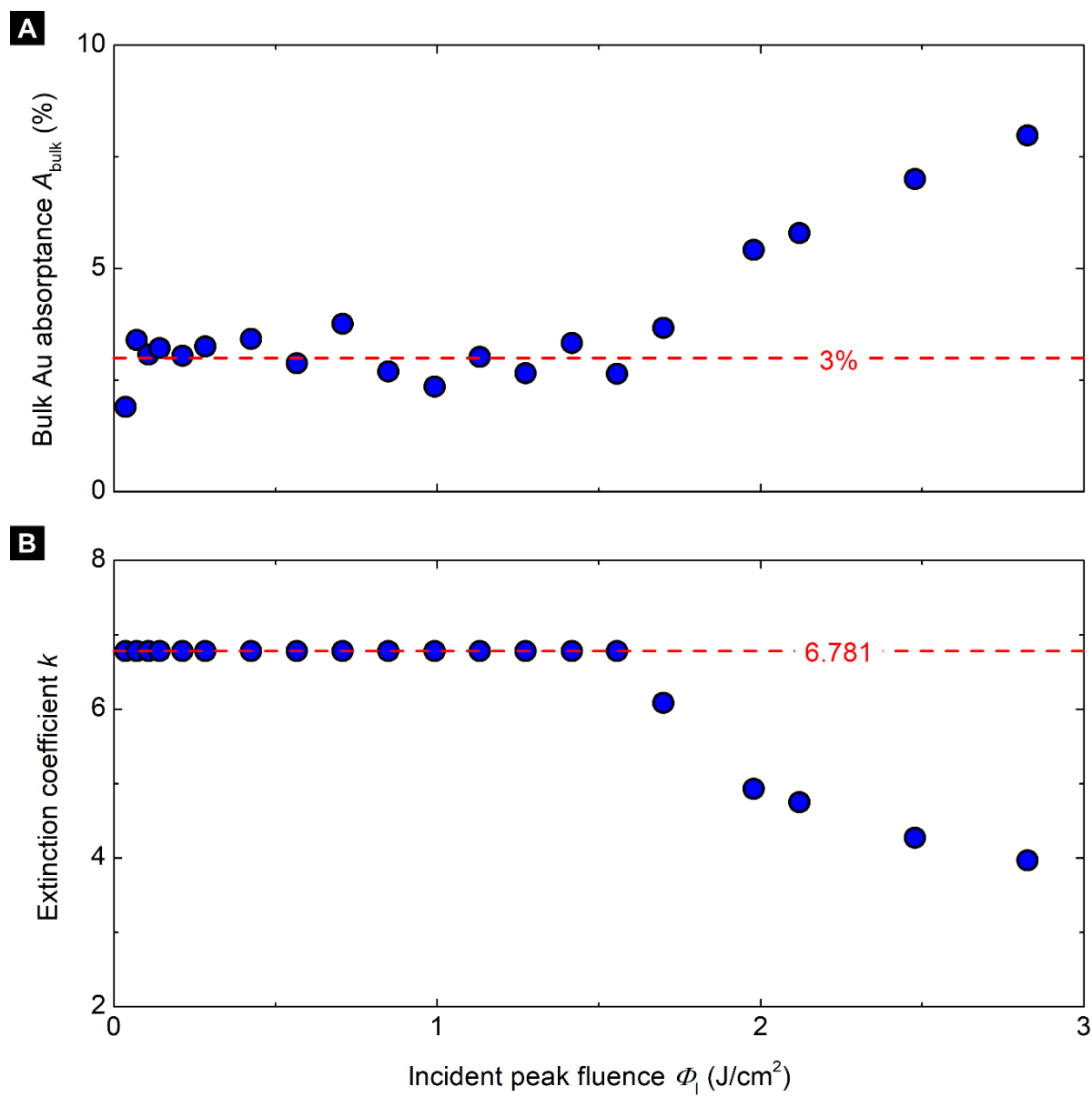

**Figure S4: A** Absorptance $A_{bulk}$ of a bulk Au target in dependence of the incident peak fluence $\Phi_l$. The target was immersed in water and irradiated under an incidence angle of 25.6° at a wavelength of 1040 nm and a pulse duration of 10 ps. The red dashed horizontal line marks the literature absorptance value of Au under these conditions. Data was extracted from [3]. **B** Calculated fluence-dependent imaginary part $k$ of the refractive index for Au. The values were derived from **A** by assuming a constant real part of the refractive index ($n = 0.247$) and calculating $k$ using Equation S2. The red dashed horizontal line marks the literature value of $k$ for Au at 1040 nm. Note that panel **A** and **B** share the same $\Phi_l$ axis.

**Table S1:** Results of Mie calculation for Au microparticles irradiated with varying peak fluence $\Phi_l$ and thus varying extinction coefficient $k$. The following quantities are shown. Absorption efficiency $Q_{abs}$, absorption cross-section $\sigma_{abs}$, scattering efficiency $Q_{sca}$, scattering cross-section $\sigma_{sca}$, extinction efficiency $Q_{ext}$ and extinction cross-section $\sigma_{ext}$.

| $\Phi_l$ (J/cm²) | $k$ | $Q_{abs}$ | $\sigma_{abs}$ (µm²) | $Q_{sca}$ | $\sigma_{sca}$ (µm²) | $Q_{ext}$ | $\sigma_{ext}$ (µm²) |
|---|---|---|---|---|---|---|---|
| ≤ 1.60 | 6.781 | 0.058 | 0.066 | 2.631 | 2.976 | 2.689 | 3.041 |
| 1.70 | 6.089 | 0.073 | 0.083 | 2.706 | 3.060 | 2.779 | 3.143 |
| 1.98 | 4.929 | 0.118 | 0.133 | 2.870 | 3.246 | 2.988 | 3.379 |
| 2.12 | 4.753 | 0.128 | 0.145 | 2.899 | 3.279 | 3.027 | 3.423 |
| 2.45 | 4.272 | 0.164 | 0.185 | 2.981 | 3.371 | 3.145 | 3.557 |
| 2.85 | 3.967 | 0.195 | 0.221 | 3.032 | 3.429 | 3.227 | 3.650 |

The absorption efficiency of the Au MPs, which for a single particle equals the MP absorptance $A_{MP}$ is shown in Figure S5.

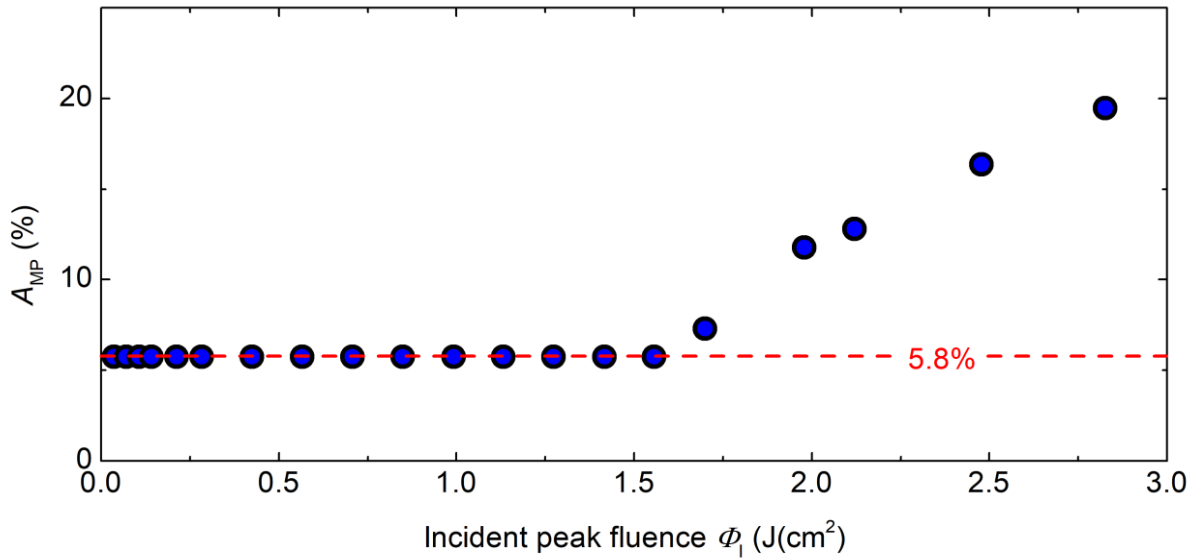

**Figure S5:** Fluence-dependent microparticle absorptance $A_{MP}$, calculated using Mie theory. The calculations were performed for spherical particles with a diameter of 1.2 µm irradiated with a wavelength of 1040 nm. The fluence-dependent $k$ from Table S1 and a constant $n = 0.245$ [4] were used as the input optical properties for Au, while for the water immersion medium the refractive index of $n = 1.324$ at 1040 nm was used [5]. The red dashed horizontal line marks the microparticle absorptance for low incident peak fluences.

## S5. Two-temperature model calculations

The electron temperature $T_e$ and lattice temperature $T_l$ following ultrashort-pulse laser irradiation are calculated based on the one-dimensional two-temperature model, given by the coupled differential Equations S3 and S4 [9].

$$c_e \cdot \frac{\partial T_e}{\partial t} = \frac{\partial}{\partial z}\left(k_e \cdot \frac{\partial T_e}{\partial z}\right) - g \cdot (T_e - T_l) + S(z,t) \tag{S3}$$

$$c_l \cdot \frac{\partial T_l}{\partial t} = g \cdot (T_e - T_l) \tag{S4}$$

Here $z$ denotes the spatial coordinate along the laser propagation direction, $t$ the time, $c_e$ the electron heat capacity, $c_l$ the lattice heat capacity, $k_e$ the electron thermal conductivity, $g$ the electron-phonon coupling factor and $S$ the laser source term.

The electron temperature dependent values of $c_e$ and $g$ for electron temperatures up to 50 kK were taken from reference [10]. The dependence of the electron thermal conductivity $k_e$ on the electron and lattice temperature was modeled by Equation S5 [11].

$$k_e(T_e, T_l) = 353\,\frac{W}{mK} \cdot \frac{(\theta_e^2 + 0.16)^{5/4} \cdot (\theta_e^2 + 0.44) \cdot \theta_e}{(\theta_e^2 + 0.092)^{1/2} \cdot (\theta_e^2 + 0.16 \cdot \theta_l)} \tag{S5}$$

In Equation S5, $\theta_e = k_B \cdot T_e / E_F$ and $\theta_l = k_B \cdot T_l / E_F$, where $E_F$ is the fermi energy.

The laser source term $S$ is given by Equation S6 and was assumed to be Gaussian in time and to decay with $z$ according to the Beer-Lambert law.

$$S(z,t) = \frac{\Phi_{abs}}{\tau_P \cdot d_{opt}} \cdot \sqrt{\frac{4\ln(2)}{\pi}} \cdot \exp\left(-\frac{z}{d_{opt}} - 4\ln(2) \cdot \left(\frac{t}{\tau_P}\right)^2\right) \tag{S6}$$

In Equation S6 $\Phi_{abs}$ denotes the absorbed peak fluence, $\tau_P = 10$ ps the FWHM laser pulse duration, and $d_{opt} = 12.2$ nm the optical penetration depth for Au at a wavelength of 1040 nm [4].

The lattice heat capacity $c_l$ was modeled by Equation S7, to account for melting [12].

$$c_l = c_{l0} + \Delta H_m \cdot \frac{1}{\sqrt{2\pi} \cdot \sigma_m} \cdot \exp\left[-\frac{1}{2} \cdot \left(\frac{T - T_m}{\sigma_m}\right)\right] \tag{S7}$$

Here $c_{l0} = 2.45 \cdot 10^6$ J/m³/K is the lattice heat capacity at room temperature, $\Delta H_m = 1.2$ J/mm³ the enthalpy of melting [13] and $T_m = 1338$ K [14] the melting temperature. The width $\sigma_m$ of the Gaussian latent heat model was chosen to be 50 K [12].

The maximum lattice temperature with respect to space and time was extracted from the two-temperature model for various absorbed peak fluences as shown in Figure S6. From this the absorbed peak fluence for the onset of phase explosion was estimated based on the criterion that the lattice temperature reaches 90% of the critical temperature [15]. Which for a critical temperature of $T_C = 7400$ K for Au [15] yielded an absorbed peak fluence of approximately 0.16 J/cm².

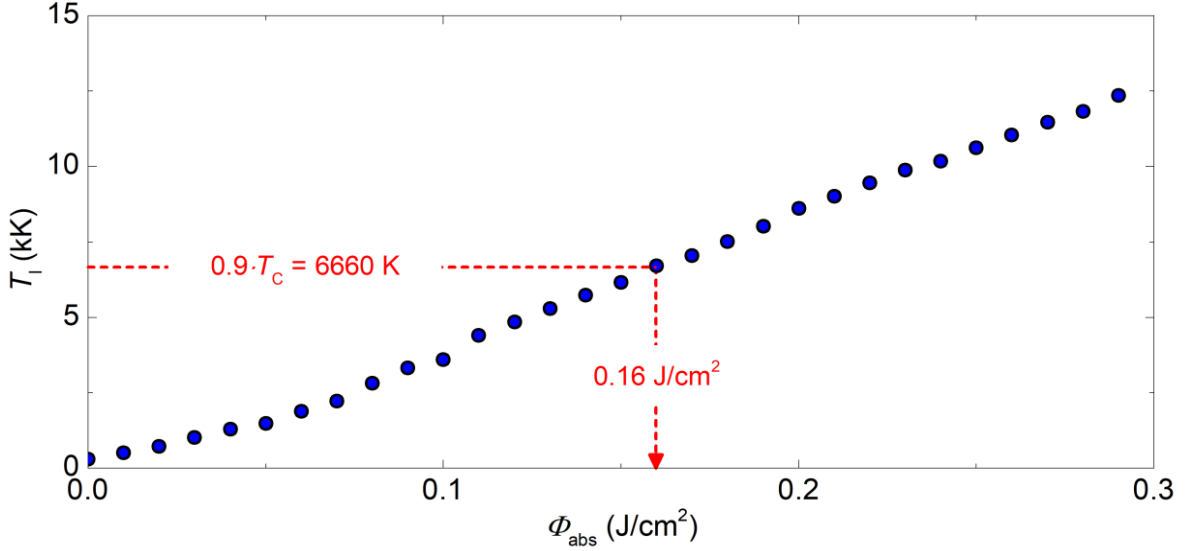

**Figure S6:** Maximum lattice temperature $T_l$ obtained by two-temperature model simulations with various absorbed peak fluences $\Phi_{abs}$. The red dashed horizontal line marks the condition for phase explosion, where the lattice temperature reaches 90% of the critical temperature. The red dashed vertical line marks the absorbed peak fluence from which the criterion for phase explosion is fulfilled.

## S6. Calculation of cavitation bubble energy from its maximal expansion

The cavitation bubble energy $E_{CB}$ is related to the maximum cavitation bubble radius $R_{max}$ by Equation S8 [16].

$$E_{CB} = \frac{4\pi}{3} \cdot (p_0 - p_v) \cdot R_{max}^3 \tag{S8}$$

Here, $p_0 = 0.1$ MPa is the ambient pressure and $p_v = 2230$ Pa the vapor pressure of water at standard conditions [16]. With a maximum cavitation bubble radius of $R_{max} = (23 \pm 1)$ µm (Main Manuscript, Figure 3a) this estimation yields a cavitation bubble energy of $E_{CB} = (5.0 \pm 0.7)$ nJ.

## S7. Calculation of cavitation bubble energy from its early expansion dynamics

The temporal evolution of the cavitation bubble radius $r_{CB}$ can be described by Equation S9, which models bubble expansion in a liquid medium [17].

$$r_{CB} = \left(\frac{25}{4\pi}\right)^{\frac{1}{5}} \cdot \left(\frac{E_{CB}}{\rho_0}\right)^{\frac{1}{5}} \cdot \Delta t^{\frac{2}{5}} \tag{S9}$$

Here, $\rho_0 = 997$ kg/m³ represents the density of water. Fitting Equation S9 to the experimental data depicted in Figure 3c of the Main Manuscript yields a bubble energy of $E_{CB} = (5.0 \pm 0.5)$ nJ.

## S8. Mie theory calculations for the probe wavelength of 520 nm

Mie theory calculations for the probe wavelength of 520 nm were carried out using a complex refractive index of N = 0.635 +i·2.072 for Au [4] and a refractive index of $n = 1.334$ for water [5]. Once more the calculation was performed for plane waves. The results of the calculation are shown in Figure S7, where the extinction efficiency $Q_{ext}$, the scattering efficiency $Q_{sca}$ and the absorption efficiency $Q_{abs}$ are plotted versus the particle diameter $D$.

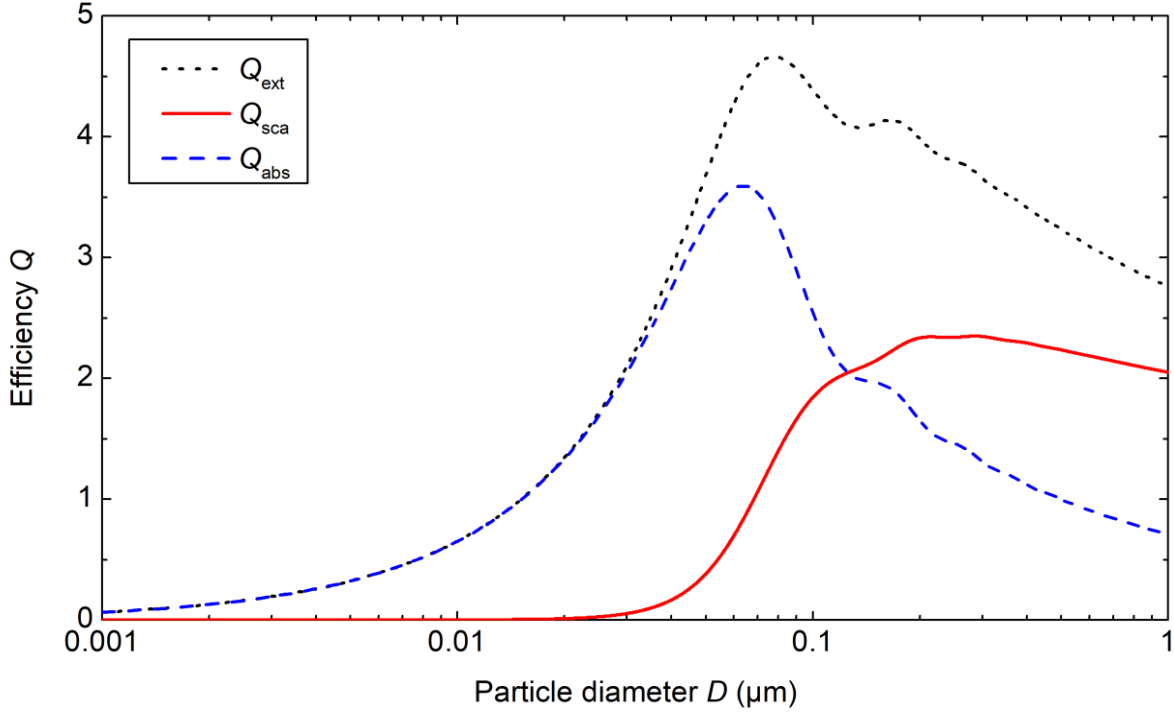

**Figure S7:** Mie theory calculations of the extinction efficiency $Q_{ext}$, scattering efficiency $Q_{sca}$ and absorption efficiency $Q_{abs}$ in dependence of the particle diameter $D$. The calculation was performed for spherical Au particles immersed in liquid and irradiated by a plane wave with a wavelength of 520 nm.

## S9. Estimation of smallest detectable nanoparticle size with the pump-probe microscopy setup

The pump-probe microscope images an area $A_{res}$ that is given by the optical resolution of the microscope onto the camera. Here $A_{res} = \pi \cdot (D_{res}/2)^2$, where $D_{res} = 0.61 \cdot \lambda / \text{NA}$ is the optical resolution of the microscope. With $\lambda = 520$ nm and NA = 0.95 this yields $D_{res} \approx 334$ nm and $A_{res} \approx 88 \cdot 10^3$ nm$^2$.

If a nanoparticle (NP) is present in the beam path, the power $P_{NP}$ removed from the probe pulse due to absorption by the NP is given by $P_{abs} = \sigma_{abs} \cdot I_0$, where $I_0$ is the probe pulse intensity and $\sigma_{abs}$ the absorption coefficient of the NP. Scattering by the NP is not considered here as it is negligible compared to absorption for Au NPs smaller than 50 nm (see Figure S8) and the aim is to find the smallest detectable NP diameter.

The transmittance of the probe beam in the presence of a NP is then given by $T = 1 - P_{abs}/P_0 = 1 - \sigma_{abs}/A_{res}$, with $P_0 = A_{res} \cdot I_0$. As the probe pulse passes through the NP, is reflected off the substrate with reflectance $R_0$ and then passes once more through the NP before being imaged onto the camera, the total reflected intensity $I_{NP}$ in the presence of a NP is given by $I_{NP} = R_0 \cdot T^2 \cdot I_0$. The relative reflectance change is then given by Equation S10.

$$\frac{\Delta R}{R_0} = \frac{R_0 \cdot T^2 \cdot I_0 - R_0 \cdot I_0}{R_0 \cdot I_0} = T^2 - 1 \tag{S10}$$

Including the transmittance $T$ in Equation S10 then yields Equation S11.

$$\frac{\Delta R}{R_0} = -2 \cdot \frac{\sigma_{abs}}{A_{res}} + \left(\frac{\sigma_{abs}}{A_{res}}\right)^2 \tag{S11}$$

Since $A_{eff} \approx 88 \cdot 10^3$ nm$^2$ is much larger than $\sigma_{abs} < 6.5 \cdot 10^3$ nm$^2$ for NPs with diameters below 50 nm (see Figure S6, where at $D = 50$ nm, $Q_{abs} = 3.3$ and thus $\sigma_{abs} = 6.5 \cdot 10^3$ nm$^2$ since $\sigma_{abs} = Q_{abs} \cdot \pi \cdot (D/2)^2$), the second term on the right side of Equation S11 can be neglected. Thus, the relative reflectance change for NPs smaller than 50 nm is given by Equation S12.

$$\frac{\Delta R}{R_0} = -2 \cdot \frac{\sigma_{abs}}{A_{res}} \tag{S12}$$

Given that the fluctuations of the $\Delta R/R_0$ signal are about 2%, Equation S12 yields the smallest detectable absorption cross-section of $\sigma_{abs} \approx 880$ nm$^2$, which corresponds to a NP diameter of approximately 26 nm (Figure S8).

## S10. Diffusion distance in dependence of nanoparticle size

The root-mean-square diffusion distance of a particle in a fluid after time $t$ is given by $l = (6 \cdot D \cdot t)^{1/2}$ where $D$ is the diffusion coefficient. According to the Stokes–Einstein equation, $D = k_B \cdot T/(6 \cdot \pi \cdot \eta \cdot r)$, the diffusion coefficient is inversely proportional to the particle radius $r$. Thus, the ratio of diffusion distances for two spherical particles of diameters $d_1$ and $d_2$ is given by Equation S13.

$$\frac{l_1}{l_2} = \sqrt{\frac{d_2}{d_1}} \tag{S13}$$

Importantly, this ratio is independent of viscosity $\eta$, temperature $T$, and diffusion time.

## S11. Literature values for single-pulse picosecond nanoparticle fragmentation thresholds

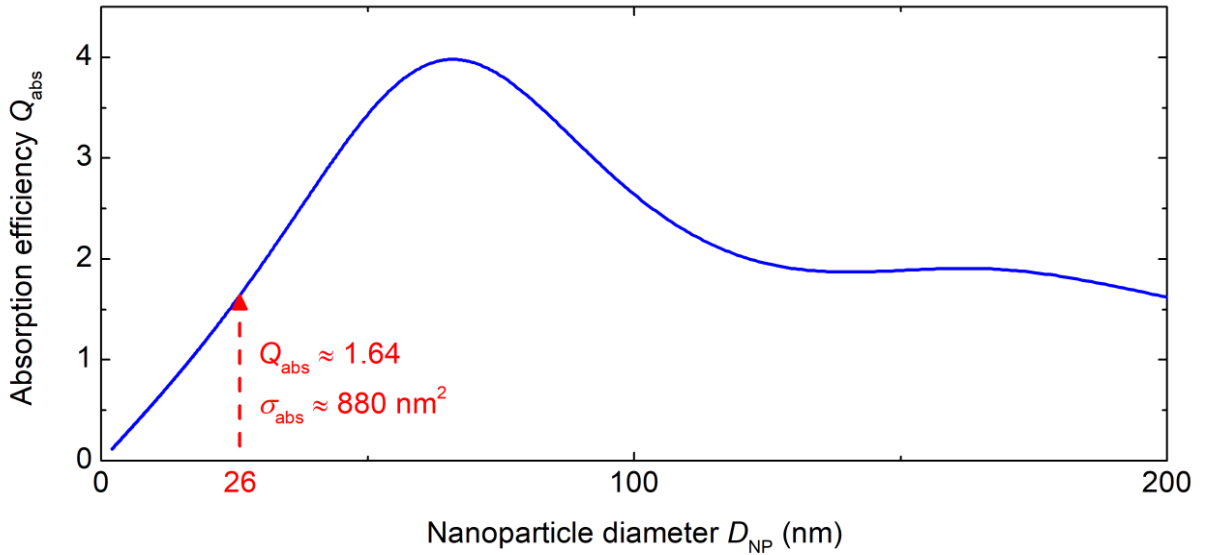

**Figure S8:** Mie theory calculations of the absorption efficiency $Q_{abs}$ as a function of nanoparticle diameter $D_{NP}$ for spherical Au nanoparticles immersed in water. The calculations were performed using MiePlot software (version 4.6.21) [8] with the assumption of a plane wave with a wavelength of 532 nm. For Au a refractive index of $n = 0.5439$ and an extinction coefficient of $k = 2.2309$ was used [4]. For water a refractive index of $n = 1.3355$ was used [5]. The red dashed horizontal arrows show Qabs and the absorption cross-section $\sigma_{abs}$ for $D_{NP} = 26$ nm (see Section S9).

**Table S2:** Literature values for nanoparticle fragmentation in liquid. The wavelength $\lambda$, the pulse duration $\tau_P$ and the nanoparticle diameter $D_{NP}$ are given for each reference. The peak fluence $\Phi_0$ marked

with † was converted from the average fluences utilizing the relation that the peak fluence is twice the average fluence [18]. The absorption efficiency $Q_{abs}$ calculated by Mie theory (see Figure S6) is given along with the calculated absorbed peak fluences $\Phi_{abs} = Q_{abs} \cdot \Phi_0$. Values marked with * denote peak fluences for complete NP fragmentation and values marked with ** denote the threshold peak fluence for fragmentation.

| Reference | $\lambda$ (nm) | $\tau_P$ (ps) | $D_{NP}$ (nm) | $\Phi_0$ (mJ/cm$^2$) | $Q_{abs}$ | $\Phi_{abs}$ (mJ/cm$^2$) |
|---|---|---|---|---|---|---|
| Ziefuss et al. [19,20] | 532 | 10 | 54 | 60[†**] | 3.12 | 187.2[*] |
| Kang et al. [21] | 532 | 28 | 15 | 1.5[**] | 0.89 | 1.3[**] |

## S12. Energy density for melting and vaporization and threshold fluence for complete microparticle melting

The energy densities $u_m$ and $u_v$ needed to melt and evaporate a volume of 1 mm$^3$ of the target were calculated by Equations S14 and S15, respectively.

$$u_m = \rho \cdot \left( \int_{T_0}^{T_m} c_p(T) dT + \Delta H_m \right) \tag{S14}$$

$$u_v = \rho \cdot \left( \int_{T_0}^{T_v} c_p(T) dT + \Delta H_m + \Delta H_v \right) \tag{S15}$$

Here $T$ denotes the temperature, $\rho$ the solids density, $T_0 = 298$ K, $T_m$ the melting point, $T_v$ the boiling point, $c_p$ the specific heat capacity at constant pressure, $\Delta H_m$ the enthalpy of fusion and $\Delta H_v$ the enthalpy of vaporization. Table S3 summarizes numerical values for the thermophysical properties used in Equations S14 and S15.

**Table S3**: Summary of thermophysical parameters for Au.

| Parameter | Value |
|---|---|
| $\rho$ (g/cm$^3$) | 19.3 [22] |
| $T_m$ (K) | 1338 [14] |
| $T_v$ (K) | 3129 [22] |
| $\rho \cdot \int_{T_0}^{T_m} c_p(T) dT$ (J/mm$^3$) | 2.6 [23,24] |
| $\rho \cdot \int_{T_0}^{T_v} c_p(T) dT$ (J/mm$^3$) | 8.7 [23,24] |
| $\rho \cdot \Delta H_m$ (J/mm$^3$) | 1.2 [13] |
| $\rho \cdot \Delta H_v$ (J/mm$^3$) | 31.8 [13] |
| $u_m$ (J/mm$^3$) | 3.8 (Equation S14) |
| $u_v$ (J/mm$^3$) | 41.7 (Equation S15) |

## S13. Estimation of the melt depth

The melt depth $d_m$ under ultrashort-pulse laser excitation at threshold (pulse duration shorter than the electron-phonon coupling time) is estimated by Equation S16, which expresses the distance hot electrons can diffuse while still transferring enough energy to raise the lattice temperature to the melting point [25,26].

$$d_{\mathrm{m}} = \left(\frac{128}{\pi}\right)^{\frac{1}{8}} \cdot \left(\frac{C_{\mathrm{l}}}{A_{\mathrm{e}} \cdot T_{\mathrm{m}}}\right)^{\frac{1}{4}} \cdot \left(\frac{\kappa_{\mathrm{e}}}{g}\right)^{\frac{1}{2}}$$ (S16)

For Au, $C_{\mathrm{l}} = 2.45 \cdot 10^6$ J/m$^3$/K [27] is the lattice heat capacity, $\gamma = 100$ J/m$^3$/K$^2$ [10] the Sommerfeld parameter, $T_{\mathrm{m}} = 1338$ K the melting temperature [14], $\kappa_{\mathrm{e}} = 317$ W/m/K [27] the electron thermal conductivity, and $g = 2.61 \cdot 10^{16}$ W/m$^3$/K [10] the electron-phonon coupling strength. Equation S16 yields a melting depth of $d_{\mathrm{m}} \approx 360$ nm.

## S14. Pressure criterion for photomechanical fracture

The threshold pressure $P_{\mathrm{th}}$ for the onset of cavitation in ultrashort pulse photomechanical fracture is given by Equation S17 [28].

$$P_{\mathrm{th}} = \sqrt{a \cdot \frac{\gamma_{\mathrm{lv}}(T)^3}{k_{\mathrm{B}} \cdot T}}$$ (S17)

Here $a \approx 0.7$ and $\gamma_{\mathrm{lv}}(T)$ is the temperature dependence of the surface tension at the liquid-vapor interface, which can be approximated by Equation S18.

$$\gamma_{\mathrm{lv}}(T) = \gamma_0 \cdot \left(1 - \frac{T}{T_C}\right)^{\mu}$$ (S18)

In Equation S19, $\gamma_0$ and $\mu$ are material dependent parameters and $T_C$ is the critical temperature, which is approximately 7400 K for Au [15]. In order to obtain the parameters $\gamma_0$ and $\mu$ for Au, $\gamma_{\mathrm{lv}}(T)$ and $\partial\gamma_{\mathrm{lv}}(T)/\partial T$ are evaluated at the melting point, which is $T_{\mathrm{m}} = 1338$ K for Au [14]. From this the following equations are obtained.

$$\gamma_{\mathrm{lv}}(T_{\mathrm{m}}) = \gamma_{\mathrm{m}} = \gamma_0 \cdot \left(1 - \frac{T}{T_C}\right)^{\mu}$$ (S19)

$$\frac{\partial\gamma_{\mathrm{lv}}(T_{\mathrm{m}})}{\partial T} = s_{\mathrm{m}} = -\frac{\mu \cdot \gamma_0}{T_C} \cdot \left(1 - \frac{T}{T_C}\right)^{\mu-1}$$ (S20)

By inserting Equation S19 into Equation S20 and solving for $\mu$, Equation S21 is obtained.

$$\mu = -\frac{s_{\mathrm{m}} \cdot T_C}{\gamma_{\mathrm{m}}} \cdot \left(1 - \frac{T_{\mathrm{m}}}{T_C}\right)$$ (S21)

With $\gamma_{\mathrm{m}} = 1.162$ N/m and $s_{\mathrm{m}} = $ -1.8$\cdot 10^{-4}$ N/m/K for Au [14], $\mu = 0.94$ is obtained. With the knowledge of $\mu$, $\gamma_0$ is determined to be 1.4 N/m from Equation S20.

Finally, by inserting the results into Equation S17, the threshold pressure of $P_{\mathrm{th}} = 7.7$ GPa is obtained for Au at the melting point.

## S15. Acoustic reflectance of gold immersed in water

The acoustic reflectance at the interface between two materials with acoustic impedance $Z_1$ and $Z_2$ is given by Equation S22.

$$R_{ac} = \frac{Z_2 - Z_1}{Z_2 + Z_1}$$ (S22)

The acoustic impedance is given by $Z = \rho \cdot c_{\mathrm{s}}$, where $\rho$ is the materials density and $c_{\mathrm{s}}$ the sound velocity of the material. Given that liquid Au has a density of 16 g/cm$^3$ and a sound velocity of 2300 m/s [29], and water has a density of 1 g/cm$^3$ and a sound velocity of 1480 m/s [30], the resulting acoustic reflectance is $R_{\mathrm{ac}} \approx 0.92$.

## S16. Fraction of scattered and absorbed pulse energy

The energy scattered and absorbed by the MP is calculated by integrating the Gaussian fluence distribution over the MP cross-section and multiplying the result by the respective efficiency, as shown in Equation S23.

$$E_i = \frac{Q_i \cdot \Phi_0 \cdot \pi \cdot w_0^2}{2} \cdot \left[ 1 - \exp\left( -\frac{D_{MP}^2}{2 \cdot w_0^2} \right) \right] \tag{S23}$$

Here, $E_i$ denotes either the scattered ($E_{sca}$) or absorbed ($E_{abs}$) pulse energy, and $Q_i$ the corresponding absorption ($Q_{abs} = 0.195$) or scattering ($Q_{sca} = 3.032$) efficiency (Section S4).

## S17. Total generated surface energy

The total generated surface energy $E_{tot}$ is calculated by Equation S24.

$$E_{tot} = \frac{\gamma}{\gamma_{det}} \cdot E_{surf} \cdot \int_0^\infty \frac{dS}{dD_F} \, dD_F \tag{S24}$$

Here $E_{surf}$ is the surface energy of Au and $\gamma_{det} = 0.015$ the mass of detected NPs and $\gamma = 0.47$ the fragmentation yield. Because the vast majority of the generated NPs are larger than 5 nm (Main Manuscript, Figure 4C), we adopt a size-independent Au NP surface energy of 1.5 J/m$^2$ consistent with atomistic calculations for nanospheres that converge to this value for diameters above 5 nm [31]. Using the total nanoparticle surface area of 1.13 μm$^2$ obtained from Figure S9, the total generated surface energy amounts to $E_{tot} \approx 0.06$ nJ.

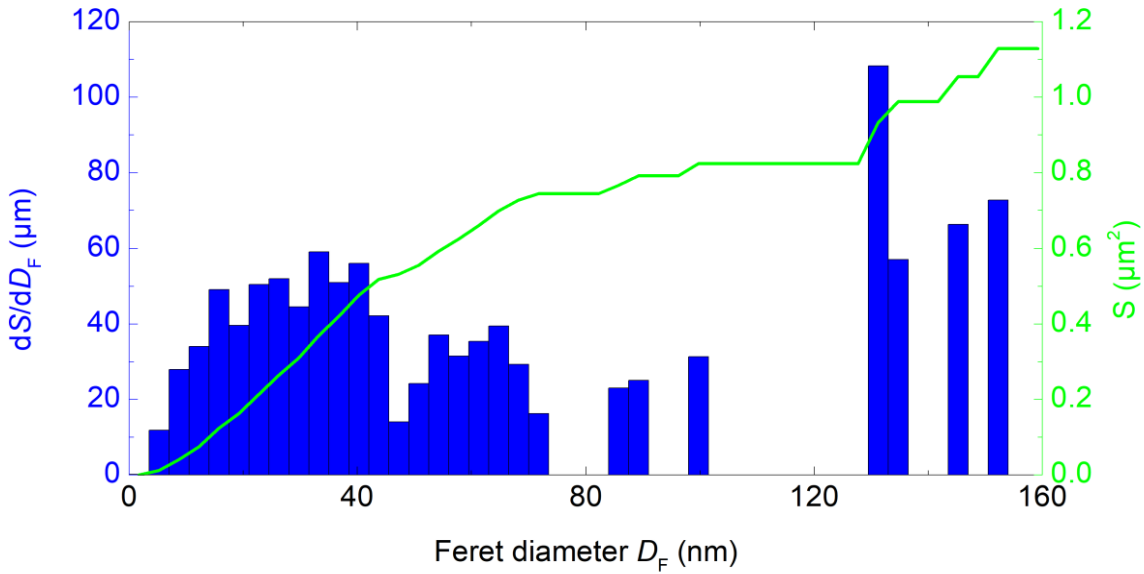

**Figure S9.** Surface-weighted nanoparticle size distribution obtained from the fragmentation of a single Au microparticle, based on analysis of 580 generated nanoparticles. The unnormalized surface-weighted distribution d$S$/d$D_F$ is shown by blue bars (left y-axis), and the cumulative surface area $S$ by the green line (right y-axis). The surface area of individual NPs is calculated assuming spherical geometry, so that the plateau of the cumulative curve directly yields the total nanoparticle surface area of 1.13 μm$^2$. This value is used in Equation S24 to determine the total generated surface energy.

## S18. Productivity estimation under state-of-the-art microparticle fragmentation conditions

To assess whether the high productivity values extrapolated in Section 4.4 of the Main Manuscript can be realized in practice, a representative experimental configuration combining the state-of-the-art ultrafast thin-disk oscillator with 550 W average power (100 µJ pulse energy at 5.5 MHz repetition rate) [32] with a state-of-the-art flat-jet reactor is considered. The flat-jet reactor, introduced for NP laser fragmentation in liquids (NP-LFL) of laser ablation in liquids generated NPs [33], has since become the state-of-the-art setup for MP fragmentation in liquids (MP-LFL) [34]. In this setup, a thin, planar liquid sheet is formed by forcing the particle suspension through a slit nozzle, producing a jet with a well-defined thickness.

To ensure stable and well-defined processing conditions in MP-LFL, it is essential that the liquid jet operates under laminar flow. Turbulent flow introduces fluctuations in jet thickness and velocity, and in extreme cases risks spray formation. With a jet velocity of 6 m/s, laminar flow has been shown to be maintained in state-of-the-art flat-jet setups [34].

To maintain single-pulse-per-volume interaction conditions, the laser repetition rate must be matched to the jet velocity. Given a jet velocity of 6 m/s [34] and a repetition rate of 5.5 MHz [32] the beam height must not exceed $d_x = 1.1$ µm to ensure that each laser pulse interacts with a fresh volume of liquid. However, such a small beam height would require very tight focusing and in addition, the Rayleigh length associated with a beam height of 1.1 µm is approximately 1 µm. This short Rayleigh length would cause dramatic fluence differences by up to a factor of 100 between the center and front/back surface of a flat jet with a thickness of 22 µm, as typically employed in state-of-the-art flat-jet setups [34]. A more reasonable beam height is estimated based on the condition that the fluence across the jet thickness must not deviate more than 1%, yielding a beam height of approximately 8 µm, which is easily obtainable with standard optics. With the given jet velocity of 6 m/s this would imply a maximum laser repetition rate of 0.75 MHz, demonstrating that the current flat-jet configuration cannot support state-of-the-art laser systems operating at 5.5 MHz while maintaining single-pulse-per-volume conditions.

Additionally, to maintain a local fluence of 2.85 J/cm$^2$ a constraint is placed upon the beam width $d_y$ perpendicular to the liquid flow direction. The spatial fluence distribution of an elliptical Gaussian beam profile is given by Equation S25.

$$\Phi(x, y) = \Phi_0 \cdot \exp\left[-2\left(\frac{x^2}{w_x^2} + \frac{y^2}{w_y^2}\right)\right] \tag{S25}$$

Here, $w_x$ and $w_y$ denote the $1/e^2$ radii along the beams minor and major axis, respectively. To ensure that the local fluence across the entire beam cross-section remains above 2.85 J/cm$^2$, the peak fluence at the beam center must be significantly higher to compensate for the steep spatial fluence gradient characteristic for Gaussian beam profiles. For a standard elliptical Gaussian beam, this corresponds to a required peak fluence of approximately $\Phi_0 = 156$ J/cm$^2$, which exceeds the edge fluence of 2.85 J/cm$^2$ at $x = w_x$ and $y = w_y$ by a factor of $e^4$ (see Equation S25). Inserting this into the standard energy-fluence relation for a Gaussian beam, $E_0 = \pi/2 \cdot w_x \cdot w_y \cdot \Phi_0$, and using the fixed beam height of $d_x = 2 \cdot w_x = 8$ µm and a pulse energy of $E_0 = 100$ µJ [32], the maximum beam width is found to be approximately $d_y = 2 \cdot w_y = 20$ µm. Given this, the resulting volumetric flow rate through the irradiated portion of the jet is approximately 9.5 mL/h. With an MP concentration of 1 g/L, which has been demonstrated to be feasible in flat-jet operation [34], and a fragmentation yield of 47%, this corresponds to a productivity of approximately 4.5 mg/h.

The productivity values derived here, based on realistic parameters of state-of-the-art flat-jet reactors and ultrafast laser systems, remain nearly five orders of magnitude below the upper productivity limit of 165 g/h (Main Manuscript Section 4.4) estimated under idealized conditions and well below the g/h limit required for MP-LFL to compete with chemical NP synthesis routes [35]. This pronounced discrepancy highlights fundamental limitations of the proposed process configuration.

## S19. Process limitations and strategic pathways for scalable microparticle fragmentation in liquids

### Flow velocity limitation and beam scanning strategy:

A major limitation of the state-of-the-art flat-jet setup is the flow velocity of 6 m/s, which under single-pulse per volume conditions limits the maximum usable laser repetition rate and thus laser power. A method to circumvent this limitation would be the employment of a scanner that translates the beam against the flow direction of the flat-jet and thus allows for higher repetition rate operation while still maintaining single-pulse per volume conditions. As in laser ablation in liquids, the minimum scanning speed must be configured in such a way as to avoid cavitation bubble shielding [36,37]. At a repetition rate of 5.5 MHz, which corresponds to a temporal inter-pulse spacing of about 180 ns the cavitation bubble has reached a radius of about 12.5 μm (Main Manuscript Figure 3C). An estimate of the required scan velocity is based on the criterion that individual irradiation events must be spaced by at least two times the cavitation bubble radius to avoid cavitation bubble shielding, yielding a required scan velocity of approximately 140 m/s. Such high scan velocities are obtainable by polygon scanners, with their feasibility for pilot-scale laser ablation in liquids already being demonstrated [36]. As the productivity increases linear with utilizable laser power and thus repetition rate, an increase of the repetition rate from 0.75 MHz to 5.5 MHz would result in a factor of 7.3 productivity increase. However, such polygon-scanners require operation at a laser duty-cycle of 50% [37], meaning that only half the available laser power can be utilized which results in a productivity increase by a factor of approximately 3.7.

### Energy distribution limitation and beam shaping strategy:

An additional major bottleneck in current MP-LFL implementations is the inefficient utilization of laser energy inherent to Gaussian beam profiles, where pulse energy is wasted because the beam center receives significantly higher fluence than required for fragmentation, while the fluence at the edges falls below the fragmentation threshold. A uniform fluence profile across the entire irradiated cross-section of the flat jet would eliminate these drawbacks. Such a profile can be achieved through spatial beam shaping using refractive or diffractive optical elements. Assuming an ideal flat-top beam profile with uniform fluence $\Phi$, the beam width can be calculated directly from the relation $\Phi = E_0/(w_x \cdot w_y)$. For a pulse energy of $E_0 = 100$ μJ, a beam height of $d_x = 2 \cdot w_x = 8$ μm, and a uniform fluence of $\Phi = 2.85$ J/cm², this yields a maximum beam width of 440 μm. The resulting volumetric flow rate through the irradiated portion of the flat jet is then 0.2 L/h, which corresponds to a productivity of 0.094 g/h for an MP concentration of 1 g/L and a fragmentation yield of 47%. This highlights that with spatial beam shaping alone, the productivity can be enhanced by a factor of more than 20 compared to the estimate in Section S18.

### Energy utilization and the critical influence of microparticle concentration:

Another major limitation in MP-LFL arises from the inefficient utilization of the available laser pulse energy, which results from the low number of MPs present within the irradiated volume at typical concentrations of 1 g/L. For the case of a uniform fluence distribution across a beam cross-section of 440 μm times 8 μm, a flat-jet thickness of 22 μm, an MP radius of 0.6 μm and an MP concentration of 1 g/L, the average number of MPs contained within the interaction volume is only approximately 5. In contrast, with an available pulse energy of 100 μJ and an estimated energy extinction of 100 nJ per fragmentation event, 1000 MPs could theoretically be fragmented per pulse. The resulting energy utilization is only about 0.5 %, calculated as the ratio of the number of MPs present within the irradiated volume to the number that could be fragmented with the available pulse energy. The remainder of the energy is transmitted through the jet as linear absorption in water is negligible for the narrow jet thickness of 22 μm.

To address this bottleneck, increasing the MP concentration is considered. The number of MPs that could maximally fill the irradiated cross-section uniformly can be estimated by dividing the beam area by the extinction cross-section of a single MP, which yields approximately 960 MPs. This corresponds to an MP concentration of 218 g/L. To put this value into context, the MPs would occupy approximately 1% of the liquid volume, based on the Au density of 19300 g/L. Since the productivity scales linearly with MP concentration in the flat jet, this increase in concentration corresponds to a 218-fold enhancement in productivity compared to a concentration of 1 g/L.

Summary of scaling factors

These scaling considerations highlight that while employing polygon scanners alone can increase productivity by a factor of 3.7, the beam shaping and concentration increase yield a far more substantial 21-fold and 218-fold improvement, respectively. This clearly identifies MP concentration as the most critical parameter for achieving high-throughput MP-LFL. However, it must be noted, that the maximum feasible concentration is yet to be experimentally validated as it was previously shown that productivity remains essentially unchanged when increasing the MP concentration beyond 3 g/L [38].

In a previous MP-LFL study on $IrO_2$ MPs [39], a fragmentation yield of 10% was reported at a comparable fluence level, approximately five times the corresponding fragmentation threshold fluence. In that study, fragmentation was induced using a Gaussian beam in a flat-jet configuration, and the reported yield most likely underestimates the achievable value, given the constraints associated with Gaussian beams as discussed in Section 4.4.3. Nevertheless, the 10% yield will serve as a conservative reference for the following analysis. Furthermore, an idealized yield of 100% is also considered as an upper bound for the productivity. To evaluate how MP-LFL upscaling performs under these assumptions, the productivity values are calculated using the different fragmentation yields. Table S4 summarizes the resulting productivity under different yield assumptions.

**Table S4:** Comparison of achievable microparticle fragmentation in liquids productivities for different experimental configurations and fragmentation yields. The "Ideal limit" refers to the maximum theoretical productivity assuming full energy utilization and a power-specific productivity of 300 mg/h/W at 550 W average power. The "State-of-the-art laser and flat jet" configuration corresponds to a 100 μJ Gaussian beam operating at a repetition rate of 0.75 MHz, focused into a flat-jet target with an MP concentration of 1 g/L, representing the state-of-the-art setup evaluated in Section S18. The subsequent rows represent this configuration enhanced by specific strategies: increased laser repetition rate through fast polygon scanners, uniform spatial energy distribution through flat-top beam shaping, increased MP concentration, and a combination of all three strategies. The numbers in parentheses indicate the corresponding multiplicative enhancement factor relative to the state-of-the-art configuration.

| Configuration | Productivity (g/h) | | |
|---|---|---|---|
| | 10% fragmentation yield (practical value) | 47% fragmentation yield (this work) | 100% fragmentation yield (maximum value) |
| Ideal limit (300 mg/h/W at 550 W) | 35 | 165 | 350 |
| State of the art laser and flat jet (Gaussian beam with 100 μJ and 0.75 MHz) | $0.96 \cdot 10^{-3}$ | $4.5 \cdot 10^{-3}$ | $9.6 \cdot 10^{-3}$ |
| + Repetition rate increase from 0.75 MHz to 5.5 MHz with polygon scanner (x 3.7) | $3.55 \cdot 10^{-3}$ | $16.7 \cdot 10^{-3}$ | $35.5 \cdot 10^{-3}$ |
| + Flat-top with 400 μm by 8 μm (x 21) | 0.02 | 0.095 | 0.2 |
| + MP concentration increased to 218 g/L (x 218) | 0.21 | 0.98 | 2.1 |
| + Polygon scanner and flat-top and MP concentration increase (x 16938) | 16.2 | 76 | 162 |

**S20. Size-distribution-induced constraints and approaches to improve nanoparticle usability**

In record-productivity laser ablation of Au in water, the sub-15 nm fraction comprised the entire detected NP population, with no particles larger than 15 nm observed [37]. In contrast, in the MP-LFL experiments presented in this study, only approximately 2% of the generated NPs fall within this size range (Main Manuscript, Figure 4), highlighting a key difference in product size distribution. Nevertheless, previous MP-LFL studies have shown that nanoclusters as small as 3 nm can be generated [38,39], demonstrating the capability of the method to produce ultrasmall particles. To increase this NP fraction and improve the competitiveness of MP-LFL, a promising strategy is the implementation of a second laser pulse that further fragments the initially generated NPs through NP-LFL [33]. The favorable power budget of MP-LFL leaves sufficient margin to carry out this secondary irradiation step and enables targeted fragmentation of the larger NP fractions (> 15 nm). To ensure effective secondary fragmentation, the timing of this second pulse must avoid shielding by the cavitation bubble generated by the primary MP-LFL event. This can be achieved in two ways: either by applying the second pulse after cavitation bubble collapse, or by delivering it before the cavitation bubble fully forms.

In the present study, cavitation bubble collapse was observed approximately 5 µs after pulse arrival (Main Manuscript, Figure 3A). Assuming a flat-jet velocity of 6 m/s, the second NP-LFL irradiation should therefore be positioned approximately 30 mm downstream from the initial MP-LFL site to avoid interaction with the generated cavitation bubbles. This demonstrates that sequential MP-LFL and NP-LFL can be implemented within a compact spatial footprint in a single flat jet.

An alternative approach is to perform NP-LFL in situ, prior to cavitation bubble formation. This has been demonstrated for laser ablation in liquids, where a second laser pulse applied 600 ps after the first led to a 9% reduction in the mass-weighted fraction of large NPs (> 15 nm), and a corresponding increase in the sub 15 nm fraction [40]. To enable in-situ NP-LFL in MP-LFL, the second laser pulse must be applied after the initial fragmentation event, but before cavitation bubble formation to avoid shielding. In this study, cavitation bubble formation was observed approximately 1 ns after pulse impact (Main Manuscript, Figure 3A), consistent with previous laser ablation in liquids results [41]. The latest time for fragmentation onset can be estimated from the time required for the pressure focusing at the MP center (see Section 4.1 of the Main Manuscript). This corresponds to the propagation time a sound wave needs to travel to the center of the MP, which, based on an average MP radius of 0.6 µm and a speed of sound in gold of 2300 m/s [29], is approximately 260 ps. Thus, in-situ NP-LFL by double-pulses or GHz bursts represents a feasible strategy for increasing the fraction of small NPs. Finally, it should be noted that under the assumption that half the available pulse energy is used for NP-LFL, both strategies for narrowing the NP size distribution would half the productivity of the initial MP-LFL process but significantly improve the product quality.